\documentclass[final,leqno,onefignum,onetabnum]{siamltex1213}
\usepackage{amsbsy,amsmath,epsfig}
\usepackage{setspace}
\title{The network picture of labor flow\thanks{This work was
supported by the Oxford Martin School under grant LC1213-006 and
INET at Oxford grant INET12-9001.}} 

\author{Eduardo L\'opez\thanks{CABDyN Complexity Centre, University of Oxford, Park End Street, Oxford, OX1 1HP, United Kingdom (\email{eduardo.lopez@sbs.ox.ac.uk}). Author supported by the James Martin 21st Century Foundation grant LC1213-006.} \and
Omar Guerrero\thanks{CABDyN Complexity Centre, University of Oxford, Park End Street, Oxford, OX1 1HP, United Kingdom (\email{omar.guerrero@sbs.ox.ac.uk}). Author supported by the Institute for New Economic Thinking grant INET12-9001.}
\and
Robert L. Axtell\thanks{Krasnow Institute for Advanced Study, George Mason University, Research I, CSC Suite, Level 3, 4400 University Drive, Fairfax, VA 22030 (\email{rax222@gmu.edu})}
}

\begin{document}
\maketitle
\slugger{mms}{xxxx}{xx}{x}{x--x}

\begin{abstract}
We construct a data-driven model of flows in graphs that captures the essential elements of the movement of workers between jobs in the companies (firms) of entire economic systems such as countries. The model is based on the observation that certain job transitions between firms are often repeated over time, showing persistent behavior, and suggesting the construction of static graphs to act as the scaffolding for job mobility. Individuals in the job market (the workforce) are modelled by a discrete-time random walk on graphs, where each individual at a node can possess two states: employed or unemployed, and the rates of becoming unemployed and of finding a new job are node dependent parameters. We calculate the steady state solution of the model and compare it to extensive micro-datasets for Mexico and Finland, comprised of hundreds of thousands of firms and individuals. We find that our model possesses the correct behavior for the numbers of employed and unemployed individuals in these countries down to the level of individual firms. Our framework opens the door to a new approach to the analysis of labor mobility at high resolution, with the tantalizing potential for the development of full forecasting methods in the future.
\end{abstract}



\pagestyle{myheadings}
\thispagestyle{plain}
\markboth{L\'OPEZ, GUERRERO, AXTELL}{THE NETWORK PICTURE OF LABOR FLOW}

\section{Introduction}
High employment is one of the central goals of any economic policy, because this is associated with economic, social and political prosperity of countries. Of the many perspectives that need to be considered to understand the problem of employment, job search has attracted a large amount of interest for its relatively well defined nature, and the perception that economic policies can have an important impact in its optimization; numerous important results have been obtained and are well summarized in reviews such as \cite{Petrongolo, Rogerson}. The main approach to understand job search is known as search and matching modeling \cite{Mortensen, Shimer}. Search and matching models broadly consist of a stochastic process by which two kinds of entities (e.g., unemployed individuals and vacancies) join to create a new match, and this joining is mediated by a success rate called the aggregate matching function~\cite{Petrongolo, Stevens}. These models have been successful at predicting quantitative and qualitative features of employment, and are well accepted~\cite{Petrongolo}.

However, despite their success, search and matching models have inherent limitations in the way they are constructed. One of those limitations, the notion of aggregation, eliminates from consideration the role played in the dynamics of employment by specific companies (firms\footnote{The definition of firm is that of enterprises with economic activity carried out by one or more persons for profit-making purposes.}) in the economy. At first glance, this may not seem critical because any country has a large number of firms, many of which are quite similar. But upon more detailed consideration, one finds it is also true that in most countries there are firms that play central roles, and when these particular firms are affected by any number of factors such as technological change, new economic policies, or competition, the impact on employment can be considerable and have downstream effects on the entire economy of the country. Indeed, empirical evidence has shown that the shocks experienced by the largest firms are responsible for the majority of fluctuations in the total production of an economy, and not necessarily because of firms' sizes, but because of the propagation effects through the entire system~\cite{Gabaix, diGiovanni}.

For some decades, governments from several countries have stored highly granular micro-data about firms and workers, constructed from social security records~\cite{Hamermesh}. However, detailed analysis of coupled firm and labor dynamics is not common in the economics literature, due to the limitations of commonly employed methods. Such data, in conjunction with the framework here proposed, offers the opportunity to uncover the specific roles that firms play in employment.

By approaching the problem of job mobility in a novel way, and using data already available, it is possible to construct more detailed models of job transitions with resolution at the level of individuals and firms in entire countries, offering a new approach to the study of labour dynamics. This is the focus of the current article. Specifically, we introduce a stochastic process on graphs that accurately represents observed employment and unemployment patterns in two comprehensive micro-datasets. Labor mobility occurs within graphs (or networks), which summarize the constraints that agents encounter while moving between jobs. These graphs are constructed as follows: vertices (nodes) represent firms, and edges represent previously observed job transitions between the firms (one or more workers changed jobs from one of the nodes to the other in a chosen time period). Workers are modeled as performing a set of simple decisions: when employed, they separate from their job with a firm-dependent probability, and when unemployed they choose to apply to one of the neighboring firms on the graph that is open to hire. These rules amount to a version of random walks on graphs (for reviews on this topic, see e.g.~\cite{Grimmett,Lovasz,Aldous}). Although the model is simple, it is able to reconstruct relevant detailed employment features of the micro-data.

One of the advantages of such a model is that it can be directly calibrated from real data down to the level of firms. This calibration, together with plausible scenarios based on the introduction of economic policies, market, or technological changes affecting particular firms, or changes in labor laws, to name a few, can potentially lead to more accurate and highly resolved forecasting of job mobility trends. Such a forecasting tool would be very valuable for those responsible for economic and labor policy-making.

As a technical consequence of our approach, we find it useful to introduce an innovative concept: firm specific unemployment. In the model, this concept is necessary because individuals that have recently stopped working at a particular firm engage in job searching only along the edges adjacent to their most recent employer, which we term \emph{local search}. Individuals then remain associated with a firm from the moment their employment finishes, to the date when they find a new job in a different firm. This concept leads to a set of new considerations about the way in which we interpret unemployment.

Finally, we present statistical evidence derived from the micro-datasets that supports our approach. First, we corroborate that our set of assumptions for the model are consistent with reality. This corroboration includes verifying that the structure of the graphs used are {\it persistent}, i.e., that job transitions over time do not simply occur randomly, but instead are regularly repeated over time, lending strength of our use of static graphs to model job movements. To our knowledge, this is the first time such a test is performed in large scale disaggregated data. With our assumptions, we show that a restricted version of our model is consistent with the general statistical features of the data including the typical number of employees at each firm, and the number of people looking for jobs after being separated from their previous firm. 

Graph approaches to the problem of job search have been considered before, albeit not with our focus. Notably, in Refs.~\cite{Boorman, Montgomery, Calvo, Schmutte,Fagiolo} and related work, job search is analyzed as a social network, where information about vacancies travels along social ties. This approach, related to the ideas and thinking of other social scientists such as Granovetter~\cite{Granovetter} have been shown to be consistent with empirical observations. However, the disadvantage of the social network framework is that social ties are not usually susceptible to the tools of economic policy, and are also hard to characterize empirically. Our approach is fundamentally different in that it focuses on the very entities in which employment takes place: firms. Somewhat related work was carried out in Ref.~\cite{Rodrigo}, where the authors consider a purely theoretical model of worker transitions between firms as a Markov process, but their approach mixes aggregate and disaggregate features, and does not tackle any empirical verification. Two of the authors of the current manuscript proposed the framework of labour flow networks in a previous publication~\cite{Guerrero}, but focused on studying their empirical properties and modeling them as the result of economic interactions. In contrast, we use the networks as a static and persistent structure that shapes labour mobility. In this publication, we attempt to build the basic modeling framework that can lead to predictions of job mobility at high resolution (down to the firm level) and show that this approach is consistent with the collected empirical evidence. 

The article is structured as follows: Sec.~\ref{sec:model} is dedicated to the construction and calculations of our labor flow model, including the derivation of the equations that broadly govern the problem, the main model predictions, and a sketch of the algorithm necessary to apply the model to data; Sec.~\ref{sec:empirics} is concerned with the empirical analysis of the data to both justify our choices in building the model, and to compare data with the predictions of the model, and; Sec.~\ref{sec:conclusions} presents the final discussion and conclusions of our work.
\section{Modelling worker movement}
\label{sec:model}
In order to provide a clear framework, we begin our detailed discussion by first introducing the assumptions of the model. This is followed by the calculation of the generating functions and moments of the distributions of numbers of employees and unemployed agents associated with a firm. We then present a treatment of the evolution of an individual agent, including job and unemployment times, which offers an alternative way to calculate the properties of the model. We finalize the section by explaining how the model can be written directly from measurable quantities in available data, and how such data needs to be used in order to predict labor flows.
\subsection{Modelling rules and assumptions}
Job search is a complicated problem, influenced by a number of factors such as skills of an agent, the type of business undertaken by a firm, effectiveness in advertising and recruiting workers by firms, etc. In order to model the search process in a tractable but efficient way, we must build a framework that is at the same time rich enough to avoid losing critical behavior, but simple enough that it can shed light on the qualitative features of the problem\footnote{Some of parameters we choose to model here, namely firm rates of acceptance of applicants, and of being open to receive applicants, are both intuitively motivated and respond to the needs of more detailed economic modeling; a complementary treatment of our model that focuses on the economic perspective is contained in an upcoming publication~\cite{AGL} by the authors.}.

Broadly speaking, there are two basic elements that need to be modelled: the structure of the economy, and the behavior of the agents.

To represent an economy, we construct a graph $G$ that encodes $N$ firms as nodes, and edges that represent allowed job transitions agents can undertake (for a review of graph theory, see~\cite{Bollobas}). The graph is assumed to be both undirected and unweighted. When dealing with real data, we develop a procedure to construct such graphs (see Sec.~\ref{sec:persistence}), but for the purpose of modelling the agent's behavior, the graph is taken as input to the model. For a theoretical investigation, one can, for instance, study job mobility in a graph sampled from an ensemble of random graphs with features relevant to a research question of interest; for a fully empirical study, one can use a single graph constructed from data for a specific economic system such as a country. Either way, we consider the graph to be static, which is to say, not changing in time. This assumption is in fact driven by our empirical findings (see Sec.~\ref{sec:empirics}). 

Firms are also characterized by a number of parameters that govern the agent dynamics, and these are also considered constant in time. One of these parameters is the probability $\lambda_i$ that an agent at firm $i$ becomes unemployed at any given time step. While the agent is employed at $i$ it is said to be in state $\mathcal{L}_i$, and if it is unemployed with $i$ being its last employer, it is in state $\mathcal{U}_i$. Probability $\lambda_i$ corresponds to the rate by which agents at $i$ move from state $\mathcal{L}_i$ to state $\mathcal{U}_i$. This probability can vary from firm to firm, but any agent employed in a specific firm has the same probability to become unemployed. An equal probability to become unemployed at a given firm is equivalent to having equal average employment time (tenure) for all employees of that firm (see Sec.~\ref{sec:tenure}).

Another parameter that is associated to a firm $i$ is the probability $v_i$ per time step that it will be accepting applications. This parameter is also assumed to be firm dependent, and it may be interpreted as a combination of a firm's financial strength, need for personnel, aggressiveness of recruitment, etc. 

The last parameter that we must define for a firm $i$ is its rate $h_i$ of hiring applicants, i.e., the probability that any individual that applies for a job at $i$ becomes employed. Parameters $h_i$ and $v_i$ play an important role in regulating the size of a firm. In real economies, even though detailed and systematic data is not available to determine $h_i$ and $v_i$, they are sensible parameters that one expects to find in associated with firms. We assume their values are in the interval $(0,1]$ in order to be meaningful in the model.

The behavior of agents is governed by the following rules. First, an agent employed at a firm (say $i$) at time step $t$ tests whether it is to remain employed (in state $\mathcal{L}_i$) or not (move to state $\mathcal{U}_i$) with probability $\lambda_i$. If it remains in $\mathcal{L}_i$, it continues onto the next time step $t+1$. If it moves to $\mathcal{U}_i$, it waits one time step and then looks for a job at step $t+1$. To search for a job, an agent in state $\mathcal{U}_i$ identifies all node neighbors $j$ that belong to $\Gamma_i$ (the node neighbors of $i$ in $G$) and that are accepting applications at that time step, each with probability $v_j$. The agent then applies to one of those neighbors with uniform probability. If none of the neighbors are open, the agent does not submit any applications and remains in $\mathcal{U}_i$ for an additional time step, when it again tries to find a job. The agent constraint of looking for a job only inside the graph leads us to define a {\it firm specific unemployment}, which reflects the continued ``association'' that agents have to their most recent employer. We assume all agents are fully aware of all neighbors that are currently accepting applications. 

With the model defined as above, we calculate analytical solutions for the average numbers of employed and unemployed agents at the firms of the graph, the probability of any agent to be employed or unemployed at a given firm, and provide the recipe for calculating other quantities of the model. Since most economies spend large proportions of time in states of small overall change, we focus on the steady state behavior of the model. This serves as a reasonable starting point for comparing the predictions of our approach to data from the real world.

Our model, at its core, corresponds to a random walk process on graphs in which some of the time scales have been modified by the waiting times that occur both in the employed and unemployed states. Formally, the process is a Markov chain, as the state of the system depends only on the previous time step.
\subsection{Evolution of the state of a firm}
To begin our detailed study, consider a given connected undirected graph $G$ with $N$ nodes, and $H$ agents distributed among the nodes of the graph (the workforce). We focus on the evolution of the system, captured by the probability distribution $Q_{i,t}(U_{i,t},L_{i,t})$ of there being $U_{i,t}$ unemployed and $L_{i,t}$ employed agents at $i$ at time $t$, where $U_{i,t},L_{i,t}$ are random variables that can take on values from 0 to $H$. To learn about the steady state of the system, we must first write down the explicit evolution equation, and consider its behavior in the steady state where $Q_{i,t}=Q^{(s)}_{i}$ for all $t$, i.e., the distribution becomes stationary in time. 

To specify the evolution equation of the system, we break down each individual mechanism of the flow process for node $i$ between time steps $t$ and $t+1$, where the number of employed and unemployed agents at $i$ and $t$ are $L_{i,t}$ and $U_{i,t}$, respectively. Consider first $\Delta_u$, the random variable that represents the number of agents becoming unemployed at a given time step. Because each agent acts independently, $\Delta_u$ has a binomial distribution, i.e.,
\begin{equation}
{\rm Pr}(\Delta_u=x|L_{i,t})={L_{i,t}\choose x}\lambda_i^{x}(1-\lambda_i)^{L_{i,t}-x}.
\label{eq:Deltau}
\end{equation}
Another mechanism affecting the number of employed agents is the acceptance or hiring rate $h_i$ of a firm. The number of new employees depends on both the number of agents that apply for a job at firm $i$, and those that are accepted. Given a number of applicants $A_{i,t}$, the probability to accept $\Delta_l$ of them is also given by a binomial
\begin{equation}
{\rm Pr}(\Delta_l=x|A_{i,t})={A_{i,t}\choose x}h_i^{x}(1-h_i)^{A_{i,t}-x}.
\label{eq:Deltal}
\end{equation}
The processes related to (\ref{eq:Deltau}) and (\ref{eq:Deltal}) are responsible for the number of agents that are employed at $i$ at time $t+1$, namely $L_{i,t+1}=L_{i,t}-\Delta_u+\Delta_l$, with probability given by the product of the two binomials above~\footnote{The distribution of $A_{i,t}$ in the steady state is the same as that for the out flow of agents from a firm, and thus it is given by (\ref{eq:Peta}) below. Here we do not make use of this result in developing the article, and thus obviate it.}.

From the standpoint of the number of unemployed agents at $t+1$, $U_{i,t+1}$ depends upon $U_{i,t}$, $\Delta_u$, and the agents in state $\mathcal{U}_i$ that find employment elsewhere, which we specify in detail below. For that purpose, we define $\gamma_{i,t}$, the subset of $\Gamma_i$ of neighbors of $i$ that are accepting job applications at time step $t$. The probability to draw any given subset $\gamma_{i,t}$ is given by the joint distribution
\begin{equation}
{\rm Pr}(\gamma_{i,t})=\prod_{j\in\gamma_{i,t}}v_j\prod_{m\in\overline{\gamma}_{i,t}}(1-v_m)
\end{equation}
where the set $\overline{\gamma}_{i,t}$ is the complement set of $\gamma_{i,t}$ with respect to $\Gamma_i$, i.e., $\gamma_{i,t}\cup\overline{\gamma}_{i,t}=\Gamma_i$ and $\gamma_{i,t}\cap\overline{\gamma}_{i,t}=\emptyset$. The use of $t$ when referring to any $\gamma_{i,t}$ is not strictly necessary, as the configurations of open neighbors are sampled independently each time step, and thus we drop reference to $t$ for these sets. When at least one neighbor is accepting applications, the probability for any agent to apply to a specific open neighbor of $i$ is equal to $1/|\gamma_{i}|$. Therefore, job applications are distributed among $\gamma_{i}$ according to a multinomial distribution. Given $U_{i,t}$ unemployed agents, with $\nu_{ij}$ applying to neighbor $j\in\gamma_{i}$, and using the symbol $\boldsymbol{\nu}_i$ to represent the entire application allocation to all nodes in $\gamma_i$, the distribution of applications to the neighbors is given by
\begin{equation}
{\rm Pr}(\boldsymbol{\nu}_i|U_{i,t},\gamma_{i})
={U_{i,t}\choose\boldsymbol{\nu}_i}\left(\frac{1}{|\gamma_{i}|}\right)^{U_{i,t}}
\end{equation}
where we have used a shorthand notation for the multinomial coefficient given by
\begin{equation}
{U_{i,t}\choose\boldsymbol{\nu}_i}={U_{i,t}\choose \nu_{ij_1},\nu_{ij_2},\dots,\nu_{ij_{|\gamma_{i}|}}}
\end{equation}
with $j_1,\dots,j_{|\gamma_{i}|}$ the elements of $\gamma_{i}$.
Given an acceptance rate of $h_{j}$ for neighbor $j$, $\eta_{ij}$ agents are hired at $j$ out of the
$\nu_{ij}$ that apply, and this random variable is also distributed in binomial fashion,
\begin{equation}
{\rm Pr}(\eta_{ij}=x|\nu_{ij})={\nu_{ij}\choose x}h_j^{x}(1-h_j)^{\nu_{ij}-x}.
\end{equation}
Altogether, representing the total accepted applications by 
$\boldsymbol{\eta}_i:=(\eta_{ij_1},\dots,\eta_{ij_{|\gamma_{i}|}})$, the probability for those acceptances is
\begin{equation}
{\rm Pr}(\boldsymbol{\eta}_i|\boldsymbol{\nu}_i,\gamma_{i})
=\prod_{j\in\gamma_{i}}{\nu_{ij}\choose\eta_{ij}}h_j^{\eta_{ij}}(1-h_j)^{\nu_{ij}-\eta_{ij}}.
\end{equation}
If in a given time step all neighbors are closed to new applicants then, by construction, $\nu_{ij}=0$ for all $j$, and similarly for $\eta_{ij}$. Symbolically, $\gamma_{i}=\emptyset$ and $\overline{\gamma}_{i}=\Gamma_i$, and this occurs with probability $\prod_{j\in\Gamma_i}(1-v_j)$. In this case, ${\rm Pr}(\boldsymbol{\nu}_i|U_{i,t},\gamma_{i}=\emptyset)$ is equal to $\delta[\boldsymbol{\nu}_i,0]$ by use of the Kronecker delta, with the convention that $\boldsymbol{\nu}_i=0$ means that all $\nu_{ij}=0$. Analogously, ${\rm Pr}(\boldsymbol{\eta}_i|\boldsymbol{\nu}_i,\gamma_{i})$ is $\delta[\boldsymbol{\eta}_i,0]$ when all neighbors are closed. Let $|\boldsymbol{\eta}_i|$ represent the total number of agents accepted into other positions, and given by $|\boldsymbol{\eta}_i|=\sum_{j\in\gamma_{i}}\eta_{ij}$. Then the number of agents in state $\mathcal{U}$ in time $t+1$ is given by $U_{i,t+1}=U_{i,t}+\Delta_u-|\boldsymbol{\eta}_i|$. In particular, when $\gamma_{i}=\emptyset$, $|\boldsymbol{\eta}_i|=0$.

To summarize the evolution, we must collect all the previous mechanisms, summing over all possible $\gamma_{i}, A_{i,t}, \Delta_u,\Delta_l,\boldsymbol{\nu}_i,\boldsymbol{\eta}_i$, and in addition, since there are multiple states at time $t$ compatible with a given state at time $t+1$, one must also sum over $U_{i,t}, L_{i,t}$. Writing a single summation symbol for the previous variables, the full expression for the evolution of $Q_{i,t}$ is given by (omitting the conditionals on the distributions)
\begin{multline}
Q_{i,t+1}(U_{i,t+1},L_{i,t+1})\\
=\sum Q_{i,t}(U_{i,t},L_{i,t})\delta[L_{i,t+1},L_{i,t}-\Delta_u+\Delta_l]
\delta[U_{i,t+1},U_{i,t}+\Delta_u-|\boldsymbol{\eta}_i|]\\
\times {\rm Pr}(\gamma_{i}){\rm Pr}(\Delta_l){\rm Pr}(A_{i,t}){\rm Pr}(\Delta_u)
\left\{\delta[\gamma_{i},\emptyset]\delta[\boldsymbol{\nu}_i,0]\delta[|\boldsymbol{\eta}_i|,0]
+(1-\delta[\gamma_{i},\emptyset]){\rm Pr}(\boldsymbol{\nu}_i){\rm Pr}(|\boldsymbol{\eta}_i|)\right\}
\label{eq:Qevol}
\end{multline}
where the $\delta[\gamma_{i},\emptyset]=1$ only when $\gamma_{i}=\emptyset$ and $0$ otherwise, and similarly $\delta[\boldsymbol{\nu}_i,0]=1$ only when all $\nu_{ij}=0$ and $0$ otherwise. The use of $|\boldsymbol{\eta}_i|$ in both terms of the brackets is a shorthand for the fact that in order to have a net outflow of agents equal to $|\boldsymbol{\eta}_i|$, one must take all possible combinations of $\{\eta_{ij}\}_{j\in\gamma_{i}}$ for given $\gamma_{i}$, and take those for which the overall flow is $|\boldsymbol{\eta}_i|$; in other words, we are implicitly using an additional factor $\delta[|\boldsymbol{\eta}_i|,\sum_{j\in\gamma_{i}}\eta_{ij}]$.

It is convenient to employ the generating function formalism~\cite{Wilf} for calculating moments of the distribution. By definition, the generating function of $Q_{i,t}(U_{i,t},L_{i,t})$ is
\begin{equation}
\mathcal{Q}_{i,t}(x,y)=\sum_{x,y}Q_{i,t}(U_{i,t},L_{i,t})x^{U_{i,t}}y^{L_{i,t}},
\end{equation}
and similarly for $Q_{i,t+1}$. Using this definition on (\ref{eq:Qevol}) applied to time $t+1$, 
one obtains the relation
\begin{multline}
\mathcal{Q}_{i,t+1}(x,y)
=\phi(1-h_i+h_i y)\left\{\mathcal{Q}_{i,t}[x,x\lambda_i+y(1-\lambda_i)]{\rm Pr}(\gamma_{i}=\emptyset)\right.\\
+\sum_{\gamma_{i}\neq\emptyset}\left.\mathcal{Q}_{i,t}\left[\langle h\rangle_{\gamma_{i}}+x(1-\langle h\rangle_{\gamma_{i}}),
x\lambda_i+y(1-\lambda_i)\right]{\rm Pr}(\gamma_{i})\right\}
\label{eq:Qgen}
\end{multline}
where $\phi$ is the generating function associated with the distribution ${\rm Pr}(A_{i,t})$, $\langle h\rangle_{\gamma_{i}}:=\sum_{j\in\gamma_{i}}h_j/|\gamma_{i}|$, and the notation $\sum_{\gamma_i\neq\emptyset}$ means that the sum runs over all possible configurations $\{\gamma_i\}$ of open neighbors of $i$ except for the case when all neighbors are not accepting applications.

The previous results can be specialized to the steady state, where 
$\mathcal{Q}_{i,t}\to\mathcal{Q}_{i}^{(s)}$ is independent of time. For now, we assume that this steady state exists and determine some of the statistical properties of the process such as the number of employed and unemployed agents; the existance of a steady state solution is shown later (Secs.~\ref{sec:avgL} and \ref{sec:agent}). 

The generating function (\ref{eq:Qgen}) can be used to calculate moments of $Q^{(s)}_i(U_i,L_i)$, although the algebra can be cumbersome for higher moments. For the average unemployment associated with firm $i$, we have
\begin{equation}
\langle U_i\rangle=\left.\frac{\partial\mathcal{Q}_i^{(s)}(x,y)}{\partial x}\right|_{x=y=1}
\label{eq:U-gen}
\end{equation} 
and for the average employment,
\begin{equation}
\langle L_i\rangle=\left.\frac{\partial\mathcal{Q}_i^{(s)}(x,y)}{\partial y}\right|_{x=y=1}.
\label{eq:L-gen}
\end{equation} 
By substituting the steady state distribution $\mathcal{Q}_i^{(s)}$ on both sides of (\ref{eq:Qgen}), and using the chain rule when taking derivatives of $x$ and $y$~\footnote{Note, for instance, that taking $x$ derivative of $\mathcal{Q}^{(s)}_i(x,x\lambda_i+y(1-\lambda_i))$ leads to $\partial\mathcal{Q}^{(s)}_i(\alpha,\beta)/\partial\alpha+\lambda_i\partial\mathcal{Q}^{(s)}_i(\alpha,\beta)/\partial\beta$ which is equal to $\langle U_i\rangle+\lambda_i\langle L_i\rangle$ when evaluated at $x=y=1$ (which leads to $\alpha=\beta=1$).}, we obtain from (\ref{eq:U-gen}) and (\ref{eq:L-gen})
\begin{equation}
\langle U_i\rangle=\frac{\lambda_i \langle L_i\rangle}
{\sum_{\gamma_{i}\neq\emptyset}\langle h\rangle_{\gamma_{i}}{\rm Pr}(\gamma_{i})}
\label{eq:Uavg}
\end{equation}
where we drop $t$ since we are in the steady state, and the sum is over all possible $\gamma_i$ except $\gamma_i=\emptyset$. Note that this expression indicates how average employment and unemployment relate to each other, but does not provide a solution that is solely based on the basic parameters of the problem. 
To construct a full solution, we must analyze in detail the flows of agents in the system, which we proceed to tackle next.
\subsection{Average employment and unemployment of a firm}
\label{sec:avgL}
In order to make progress, we study the full distribution ${\rm Pr}(|\boldsymbol{\eta}_i|)$ of outgoing agents from firm $i$. Let us recall that the distribution of outgoing application allocations is governed by ${\rm Pr}(\boldsymbol{\nu}_i|U_{i,t},\gamma_{i})$ and the hirings by ${\rm Pr}(\boldsymbol{\eta}_i|\boldsymbol{\nu}_i,\gamma_{i})$. Furthermore, the overall flow is also dependent on $\gamma_{i}$ and $U_{i,t}$ (through $Q_{i,t}(U_{i,t},L_{i,t})$). We must also keep in mind that $\gamma_{i}$ can be the empty set when no neighbors are receiving applicants. Therefore, summing over $U_{i,t},L_{i,t}, \boldsymbol{\nu}_i$ and $\{\gamma_i\}$ (the set of all possible configurations of open and closed neighbors to $i$), we have
\begin{multline}
{\rm Pr}(|\boldsymbol{\eta}_i|)
=\sum Q_{i,t}(U_{i,t},L_{i,t}){\rm Pr}(\gamma_{i})
\{\delta[\gamma_{i},\emptyset]\delta[\boldsymbol{\nu}_i,0]\delta[|\boldsymbol{\eta}_i|,0]\\
+(1-\delta[\gamma_{i},\emptyset]){\rm Pr}(\boldsymbol{\nu}_i|U_{i,t},\gamma_i){\rm Pr}(\boldsymbol{\eta}_i|\boldsymbol{\nu}_i,\gamma_i)\},
\label{eq:Peta}
\end{multline}
where we have kept the conditionals to avoid confusion. The corresponding generating function for ${\rm Pr}(|\boldsymbol{\eta}_i|)$ is given by
\begin{equation}
\psi(x)={\rm Pr}(\gamma_{i}=\emptyset)
+\sum_{\gamma_{i}\neq\emptyset}\sum_{L_{i,t}}\mathcal{Q}_{i,t}\left[1-\langle h\rangle_{\gamma_{i}}
+x\langle h\rangle_{\gamma_{i}},L_{i,t}\right]{\rm Pr}(\gamma_{i,t}).
\end{equation}
Since $\psi(x)=\sum {\rm Pr}(|\boldsymbol{\eta}_i|)x^{|\boldsymbol{\eta}_i|}$, the sum over $L_{i,t}$ remains expressed since there is no additional variable $y$ that sums over the second argument of $Q_{i,t}(U_{i,t},L_{i,t})$. Despite this, we still use $\mathcal{Q}_{i,t}$ to represent the generating function summed only over $U_{i,t}$. In the steady state, the average outflow is given by the first derivative $d\psi/dx$ evaluated at $x=1$, which produces
\begin{equation}
\langle|\boldsymbol{\eta}_i|\rangle=\langle U_i\rangle\sum_{\gamma_i\neq\emptyset}\langle h\rangle_{\gamma_i}{\rm Pr}(\gamma_i)
\label{eq:etaavg1}
\end{equation}
and with the use of (\ref{eq:Uavg}), 
\begin{equation}
\langle|\boldsymbol{\eta}_i|\rangle=\lambda_i\langle L_i\rangle
\label{eq:etaavg2}
\end{equation}
which is intuitively sound, as the number of agents that become unemployed and look for jobs
is on average $\lambda_i\langle L_i\rangle$ and therefore they must flow elsewhere for the steady state to
be achieved.
A similar calculation leads to the average steady state agent flow along a particular edge, which is
\begin{equation}
\langle \eta_{ij}\rangle=\langle U_i\rangle h_j\sum_{\{\gamma_{i}^{(j)}\}}\frac{1}{|\gamma_i^{(j)}|}{\rm Pr}(\gamma_i^{(j)})
\label{eq:etaijavg}
\end{equation}
where $\{\gamma_{i}^{(j)}\}$ is the set of all possible configurations of open and closed neighbors of $i$ in which node $j$ is guaranteed to be present (open), and the sum is over all such configurations.

The steady state condition is satisfied if the average flows into and out of a node (firm) are equal. This implies
\begin{equation}
\langle |\boldsymbol{\eta}_i|\rangle=\langle\Delta_l\rangle=\sum_{j\in\Gamma_i}\langle \eta_{ji}\rangle.
\end{equation}
Using (\ref{eq:Uavg}), (\ref{eq:etaavg2}), and (\ref{eq:etaijavg}), one can restate this as
\begin{equation}
\lambda_i\langle L_i\rangle=\sum_{j\in\Gamma_i}
\frac{\lambda_jh_i\langle L_j\rangle\sum_{\{\gamma_j^{(i)}\}}{\rm Pr}(\gamma_j^{(i)})/|\gamma_j^{(i)}|}
{\sum_{\gamma_j\neq\emptyset}\langle h\rangle_{\gamma_j}{\rm Pr}(\gamma_j)}.
\label{eq:Lsys1}
\end{equation}
This expression provides a system of equations that can in principle be solved for all $\langle L_i\rangle$, provided such solution exists. 

To understand this further, we write (\ref{eq:Lsys1}) in matrix form making use of the adjacency matrix of the graph, $\mathbf{A}$, for which $\mathbf{A}_{ij}=\mathbf{A}_{ji}=1$ if $i$ and $j$ have an edge connecting them, and zero otherwise. This produces the expression
\begin{equation}
\sum_{j=1}^N\left[\mathbf{A}_{ij}\frac{h_i\sum_{\{\gamma_j^{(i)}\}}{\rm Pr}(\gamma_j^{(i)})/|\gamma_j^{(i)}|}
{\sum_{\gamma_j\neq\emptyset}\langle h\rangle_{\gamma_j}{\rm Pr}(\gamma_j)}-\delta[i,j]\right]\lambda_j\langle L_j\rangle=0
\end{equation}
for all $i$. This represents a homogeneous system of linear equations, which always has the trivial null solution,
and has non-trivial solutions if and only if the matrix contained inside brackets is singular which,
among other things, implies that the matrix does not have full rank~\cite{Hoffman}. 
To show that our model has non-trivial solutions indeed, we define the matrix $\boldsymbol{\Lambda}$, 
with element $\boldsymbol{\Lambda}_{ij}$ corresponding to the expression inside brackets
\begin{equation}
\boldsymbol{\Lambda}_{ij}:=\mathbf{A}_{ij}
\frac{h_i\sum_{\{\gamma_j^{(i)}\}}{\rm Pr}(\gamma_j^{(i)})/|\gamma_j^{(i)}|}
{\sum_{\gamma_j\neq\emptyset}\langle h\rangle_{\gamma_j}{\rm Pr}(\gamma_j)}-\delta[i,j].
\label{eq:Lambda}
\end{equation}
This matrix does not possess full rank as can be explicitly seen from the fact that all columns
add to zero. To show this, we first sum $\boldsymbol{\Lambda}_{ij}$ over $i$
\begin{equation}
\sum_{i=1}^N\boldsymbol{\Lambda}_{ij}=-1+\sum_{i=1}^N\mathbf{A}_{ij}
\frac{h_i\sum_{\{\gamma_j^{(i)}\}}{\rm Pr}(\gamma_j^{(i)})/|\gamma_j^{(i)}|}
{\sum_{\gamma_j\neq\emptyset}\langle h\rangle_{\gamma_j}{\rm Pr}(\gamma_j)}
\label{eq:Lambdasum}
\end{equation}
where $-1$ comes from $-\sum_i\delta[i,j]$. We can now show that the numerator and denominator of the second term are indeed equal. To see this in detail, we organize the elements of $\{\gamma_j^{(i)}\}$ by cardinality $|\gamma_j^{(i)}|$, and rewrite the numerator as
\begin{equation}
\sum_{i=1}^N\mathbf{A}_{ij}h_i\sum_{\{\gamma_j^{(i)}\}}{\rm Pr}(\gamma_j^{(i)})/|\gamma_j^{(i)}|
=\sum_{c=1}^{|\Gamma_j|}\frac{1}{c}\sum_i \mathbf{A}_{ij}h_i\sum_{|\gamma_j^{(i)}|=c}{\rm Pr}(\gamma_j^{(i)}),
\end{equation}
where the last sum is over all elements of $\{\gamma_j^{(i)}\}$ with equal size $c$. Now, the sum over $i$ guarantees
that each neighbor of $j$ belonging to a particular $\gamma_j^{(i)}$ is summed, along with the corresponding $h_r$,
where $r\in\gamma_j^{(i)}$. Therefore, the sum over $i$ can be rewritten as
\begin{equation}
\sum_i \mathbf{A}_{ij}h_i\sum_{|\gamma_j^{(i)}|=c}{\rm Pr}(\gamma_j^{(i)})
=\sum_{|\gamma_j|=c}\left(\sum_{r\in\gamma_j} h_r\right){\rm Pr}(\gamma_j)
\end{equation}
and inserting this into the sum over $c$ leads to
\begin{equation}
\sum_{c=1}^{|\Gamma_j|}\frac{1}{c}\sum_{|\gamma_j|=c}\left(\sum_{r\in\gamma_j} h_r\right){\rm Pr}(\gamma_j)
=\sum_{\gamma_j\neq\emptyset}\frac{\sum_{r\in\gamma_j}h_r}{|\gamma_j|}{\rm Pr}(\gamma_j)
=\sum_{\gamma_j\neq\emptyset}\langle h\rangle_{\gamma_j}{\rm Pr}(\gamma_j)
\end{equation}
Therefore,
\begin{equation}
\sum_{i=1}^N\mathbf{A}_{ij}h_i\sum_{\{\gamma_j^{(i)}\}}{\rm Pr}(\gamma_j^{(i)})/|\gamma_j^{(i)}|
=\sum_{\gamma_j\neq\emptyset}\langle h\rangle_{\gamma_j}{\rm Pr}(\gamma_j)
\end{equation}
which means that for all $j$, (\ref{eq:Lambdasum}) is identically zero. 

The fact that $\boldsymbol{\Lambda}$ has reduced rank can also be seen from (\ref{eq:etaavg1}) and (\ref{eq:etaijavg}), which imply
\begin{equation}
\boldsymbol{\Lambda}_{ij}=\mathbf{A}_{ij}\frac{\langle\eta_{ji}\rangle}{\langle\boldsymbol{|\eta|}_j\rangle}-\delta[i,j],
\label{eq:lambda-flows}
\end{equation}
and because, by definition $|\boldsymbol{\eta}_j|=\sum_{j}\eta_{ji}$, one arrives at $\sum_{i=1}^N\boldsymbol{\Lambda}_{ij}=0$ as before. But matrix $\boldsymbol{\Lambda}$, as expressed in (\ref{eq:lambda-flows}), manifestly represents the Laplacian matrix of a random walk with heterogeneous transitions probabilities on the edges of the graph, a well-understood process~\cite{Bollobas}. Such walks are known to be ergodic, and their convergence rate can be calculated through the spectral properties of $G$.

To develop the rest of the theory, we focus on graphs with a single connected component containing all nodes $N$ (a connected graph), and explain the more general case below (see Sec.~\ref{sec:algorithm}). Defining the column matrix
\begin{equation}
\mathbf{X}_j=\lambda_j\langle L_j\rangle
\end{equation}
for the average employment in the firms of the system, one obtains the homogeneous system of equations
\begin{equation}
\boldsymbol{\Lambda}\mathbf{X}=\mathbf{0},
\label{eq:LambdaEq}
\end{equation}
where the right hand side is the column matrix of dimension $N\times 1$ of zeros. The non-trivial solutions to this system, if they exist, depend on $\boldsymbol{\Lambda}$ being singular, which is valid in our case. Since the matrix for a connected graph has rank $N-1$, its kernel is one-dimensional, and thus, to choose a unique solution that belongs to the kernel of $\boldsymbol{\Lambda}$ one needs a single additional condition. In our case, this condition corresponds to the total number of agents $H$ in the system, i.e.
\begin{equation}
\sum_{i=1}^N(\langle L_i\rangle+\langle U_i\rangle)=H.
\label{eq:cond}
\end{equation}
Application of (\ref{eq:cond}), as illustrated below, leads to the desired unique solution.

Solving (\ref{eq:LambdaEq}) and (\ref{eq:cond}) in the general case does not produce compact solutions. However, it is possible to obtain some explicit solutions for simple cases, such as when the probability that a firm is open to hire is homogeneous over all nodes ($v_j=v$ for all $j$). Explicitly, note that in the homogeneous case ${\rm Pr}(\gamma_j)\rightarrow v^{|\gamma_j|}(1-v)^{|\Gamma_j|-|\gamma_j|}$. It is common in the networks and graph theory literature to use the notation $k_j=|\Gamma_j|$, and refer to $k_j$ as the degree of node $j$. Then,
\begin{equation}
\sum_{\{\gamma_j^{(i)}\}}{\rm Pr}(\gamma_j^{(i)})/|\gamma_j^{(i)}|\rightarrow
\sum_{|\gamma_j^{(i)}|=1}^{k_j}{k_j-1\choose |\gamma_j^{(i)}|-1}\frac{v^{|\gamma_j^{(i)}|}(1-v)^{k_j-|\gamma_j^{(i)}|}}
{|\gamma_j^{(i)}|}
=\frac{1-(1-v)^{k_j}}{k_j}.
\end{equation}
For the sum $\sum_{\gamma_j\neq\emptyset}\langle h\rangle_{\gamma_j}{\rm Pr}(\gamma_j)$, we note that
each acceptance rate $h_i$ for $i\in\Gamma_j$ appears ${k_j-1\choose|\gamma_j|-1}$ times among all the terms 
where there are $|\gamma_j|$ open neighbors to $j$. One can then write in the homogeneous case
\begin{equation}
\sum_{\gamma_j\neq\emptyset}\langle h\rangle_{\gamma_j}{\rm Pr}(\gamma_j)\rightarrow
\sum_{|\gamma_j|=1}^{k_j}{k_j-1\choose|\gamma_j|-1}\frac{\sum_{i\in\Gamma_j} h_i}{|\gamma_j|}
v^{|\gamma_j|}(1-v)^{k_j-|\gamma_j|}
=\langle h\rangle_{\Gamma_j}(1-(1-v)^{k_j}),
\end{equation}
where $\langle h\rangle_{\Gamma_j}:=\sum_{i\in\Gamma_j}h_i/k_j$, i.e., the average hiring
rate of the full neighbor set of $j$.
In this case, the matrix $\boldsymbol{\Lambda}$ takes on the form
\begin{equation}
\boldsymbol{\Lambda}^{(v)}_{ij}=\frac{\mathbf{A}_{ij}h_i}{k_j\langle h\rangle_{\Gamma_j}}-\delta[i,j],
\end{equation}
and in the very simple example where all $h_i$ are equal, $\boldsymbol{\Lambda}$ is equal to the usual normalized
Laplacian for random walks on an unweighted graph. To refer to this model, we introduce the superscript $(v)$ as a reminder that this quantity is now constant. By inspection, we can find a solution for $\mathbf{X}$, which provides
\begin{equation}
\langle L_i\rangle^{(v)}=\frac{\rho h_i\langle h\rangle_{\Gamma_i}k_i}{\lambda_i},
\label{eq:Lh}
\end{equation}
and
\begin{equation}
\langle U_i\rangle^{(v)}=\frac{\rho h_ik_i}
{1-(1-v)^{k_i}},
\label{eq:Uh}
\end{equation}
where $\rho$ is a constant that can be obtained by imposing (\ref{eq:cond}), and is given by
\begin{equation}
\rho
=\frac{H}{\sum_{i\in G}h_i\langle h\rangle_{\Gamma_i}k_i\left[\frac{1}{\lambda_i}
+\frac{1}{\langle h\rangle_{\Gamma_i}[1-(1-v)^{k_i}]}\right]}.
\label{eq:rho}
\end{equation}
This quantity has an intuitive interpretation, in that it captures the average flow rate of workers all through the system. 
\subsection{The agent perspective}
\label{sec:agent}
The results from the previous section were derived with the population of agents in mind. In that context we showed that there are non-trivial solutions for $\langle L_i\rangle$ in the model, and derived the equation that describes the system. 

An alternative approach to the solution of the model is to consider the single agent perspective. This approach is a valid alternative to solve the model because agents are non-interacting, and therefore the dynamics of any one of them are sufficient to rederive the results above. In this section, we elaborate on this approach still within the context of connected graphs.

Taking the view of an individual agent, it is convenient to define the probabilities $r(i,t)$ and $s(i,t)$ that the agent
would be, respectively, employed or unemployed at the node $i$ at time $t$. These two probabilities, explained in detail below, satisfy the equations
\begin{eqnarray}
r(i,t)&=&(1-\lambda_i)r(i,t-1)
+h_i\sum_{j\in\Gamma_i}s(j,t-1)\sum_{\{\gamma^{(i)}_j\}}\frac{1}{|\gamma^{(i)}_j|}{\rm Pr}(\gamma^{(i)}_j)\\
s(i,t)&=&\lambda_i r(i,t-1)
+s(i,t-1)\left[\sum_{\gamma_i\neq\emptyset}{\rm Pr}(\gamma_i)\frac{1}{|\gamma_i|}\sum_{j\in\gamma_i}(1-h_j)
+{\rm Pr}(\emptyset)\right],
\end{eqnarray}
where the square brackets of the second equation can be simplified to $1-\sum_{\gamma_i\neq\emptyset}\langle h\rangle_{\gamma_i}{\rm Pr}(\gamma_i)$. The first equation states that the probability for an agent to be at node $i$ at time $t$ is given by the probability to be at node $i$ at time $t-1$ and not become unemployed, plus the probability that the agent is unemployed at one of the neighbors of $i$, that $i$ is accepting applications, that the agent choses to apply to $i$, and that the application by the agent leads to being hired. The second equation states that the probability to be unemployed at $i$ at time $t$ is given by the probability to be employed at $i$ at time $t-1$ and be separated with probability $\lambda_i$, or to have been unemployed at time $t-1$ at $i$ but not find a job among the neighbors of $i$, either because they are all closed, or because the agent chooses to apply to one of the neighbors and is not hired.

The previous results lead to a set of difference equations that can be written as a matrix equation with block structure. In the steady state, this matrix equation is simplified because the conditions $r(i,t)-r(i,t-1)=0$ and $s(i,t)-s(i,t-1)=0$ are satisfied. Given that in the steady state $r$ and $s$ no longer depend on time, we write the equations for $r(i,t)\to r_{\infty}(i)$ and $s(i,t)\to s_{\infty}(i)$ in the steady state
\begin{eqnarray}
0&=&-\lambda_i r_\infty(i)
+h_i\sum_{j\in\Gamma_i}s_\infty(j)\sum_{\{\gamma^{(i)}_j\}}\frac{1}{|\gamma^{(i)}_j|}{\rm Pr}(\gamma^{(i)}_j)\label{eq:rsteady}\\
0&=&\lambda_i r_\infty(i)-s_\infty(i)\sum_{\gamma_i\neq\emptyset}\langle h\rangle_{\gamma_i}{\rm Pr}(\gamma_i),\label{eq:ssteady}
\end{eqnarray}
which can be solved by first expressing $s_{\infty}(i)$ in terms of $r_{\infty}(i)$ 
\begin{equation}
s_\infty(i)=r_\infty(i)\frac{\lambda_i}{\sum_{\gamma_i\neq\emptyset}\langle h\rangle_{\gamma_i}{\rm Pr}(\gamma_i)},
\end{equation}
and substituting into (\ref{eq:rsteady}) to produce 
\begin{equation}
\sum_{j=1}^N\left[\frac{\mathbf{A}_{ij}h_i\sum_{\{\gamma^{(i)}_j\}}{\rm Pr}(\gamma^{(i)}_j)/|\gamma^{(i)}_j|}
{\sum_{\gamma_j\neq\emptyset}\langle h\rangle_{\gamma_j}{\rm Pr}(\gamma_j)}-\delta[i,j]\right]\lambda_j r_\infty(j)=0.
\end{equation}
The matrix in brackets is simply $\boldsymbol{\Lambda}$ defined in (\ref{eq:Lambda}). As we have seen, the matrix does not have complete rank, guaranteeing the existence of non-trivial solutions. The steady state with homogeneous probability $v_i=v$ for firms to be open leads to solutions similar as those above for the entire population of agents, but with a different $\rho$ which we relabel as $\chi$, i.e.,
\begin{eqnarray}
r_{\infty}^{(v)}(i)&=&\frac{\chi h_i\langle h\rangle_{\Gamma_i}k_i}{\lambda_i}\label{eq:rinf}\\
s_{\infty}^{(v)}(i)&=&\frac{\chi h_ik_i}{1-(1-v)^{k_i}}\label{eq:sinf}\\
\chi&=&\rho(H=1)=\frac{1}{\sum_{i\in G} h_i\langle h\rangle_{\Gamma_i}k_i\left[\frac{1}{\lambda_i}
+\frac{1}{\langle h\rangle_{\Gamma_i}[1-(1-v)^{k_i}]}\right]},\label{eq:chi}
\end{eqnarray}
where the normalization condition is $\sum_i [r(i)+s(i)]=1$ (independent of the steady state or the condition $v_i=v$). 

Once $r_{\infty}(i)$ and $s_{\infty}(i)$ have been determined for the model of interest (homogeneous or heterogeneous $h, v$, etc.), the number of employed and unemployed agents at firm $i$ can then be computed via
\begin{equation}
{\rm Pr}(L_i)={H\choose L_i}[r_{\infty}(i)]^{L_i}[1-r_{\infty}(i)]^{H-L_i}
\label{eq:agent-L}
\end{equation}
and 
\begin{equation}
{\rm Pr}(U_i)={H\choose U_i}[s_{\infty}(i)]^{U_i}[1-s_{\infty}(i)]^{H-U_i}.
\label{eq:agent-U}
\end{equation}
These expressions reproduce the results presented in the previous sections, and can also be used to calculate higher moments of the distributions on the basis of the steady state distributions for a single agent. For instance, the variance for $L_i$ and $U_i$ can be calculated via well-known expressions for binomial distributions, yielding 
\begin{equation}
{\rm var}(L_i)=\langle L_i^2\rangle-\langle L_i\rangle^2
=Hr_{\infty}(i)[1-r_{\infty}(i)]
\label{eq:agent-varL}
\end{equation}
and 
\begin{equation}
{\rm var}(U_i)=\langle U_i^2\rangle-\langle U_i\rangle^2
=Hs_{\infty}(i)[1-s_{\infty}(i)].
\label{eq:agent-varU}
\end{equation}
From the practical standpoint, it is useful to realize that, if $r_{\infty}(i),s_{\infty}(i)$ need to be estimated (say numerically), (\ref{eq:agent-L}), (\ref{eq:agent-U}), (\ref{eq:agent-varL}), (\ref{eq:agent-varU}), and other quantities that can be calculated as functions of $r_{\infty}(i),s_{\infty}(i)$ become particularly useful because it is no longer necessary to try to solve (\ref{eq:LambdaEq}) and (\ref{eq:cond}) directly, which could be demanding for very large economies. Instead, estimates of $r_{\infty}(i),s_{\infty}(i)$ could be utilized to arrive at meaningful results.
\subsection{Employment tenure and unemployment spells}
\label{sec:tenure}
In our model, the mechanism for job separation is characterized by a geometric distribution. Hence, an agent employed in firm $i$ has a probability $\lambda_i$ to be separated per time step. Therefore, the distribution ${\rm Pr}(t^{(l)})$ of employment duration $t^{(l)}$ (also known as job tenure), is given by
\begin{equation}
{\rm Pr}(t^{(l)})=(1-\lambda_i)^{t^{(l)}-1}\lambda_i.
\label{eq:Lspell}
\end{equation}
The average time of employment in firm $i$ is given by
\begin{equation}
\langle t^{(l)}_{i}\rangle=\frac{1}{\lambda_i}.
\end{equation}

A similar calculation provides us with the duration $t^{(u)}$ of unemployment spells. In particular, the probability for an agent to find a job among the neighbors of firm $i$ (i.e., the effective rate of hiring) depends on $\xi_i:=\sum_{\gamma_i}\langle h\rangle_{\gamma_i}{\rm Pr}(\gamma_i)$, and therefore, the distribution of unemployment spells is given by
\begin{equation}
{\rm Pr}(t^{(u)})=(1-\xi_i)^{t^{(u)}-1}\xi_i
\label{eq:Uspell}
\end{equation}
with average unemployment duration 
\begin{equation}
\langle t^{(u)}_i\rangle=\frac{1}{\xi_i}.
\end{equation}
In the case of homogeneous probability for firms to accept applications ($v_i=v$ for all $i$), unemployment spells
are charaterized by $\xi^{(v)}_i=\langle h\rangle_{\Gamma_i}[1-(1-v)^{k_i}]$. This allows us to rewrite (\ref{eq:Uh}) and (\ref{eq:sinf}) in the more intuitive forms
\begin{equation}
\langle U_i\rangle^{(v)}=\frac{\rho h_ik_i}
{1-(1-v)^{k_i}}=\frac{\rho h_i\langle h\rangle_{\Gamma_i}k_i}{\xi^{(v)}_i}
\label{eq:Uh-simpler}
\end{equation}
and
\begin{equation}
s_{\infty}^{(v)}(i)=\frac{\chi h_ik_i}{1-(1-v)^{k_i}}=\frac{\chi h_i\langle h\rangle_{\Gamma_i}k_i}{\xi^{(v)}_i}
\end{equation}
which also exposes the symmetrical nature of the values for average employment and unemployment as we see in detail in (\ref{eq:LoUvsXioLamb}) below.

The characteristic times calculated above help us provide an intuitive understanding of (\ref{eq:rho}) and (\ref{eq:chi}). Specifically, note that the joint distribution of being employed for $t^{(l)}$ time steps and subsequently unemployed for $t^{(u)}$ time steps is distributed as the convolution of the two geometric distributions above, and the average time for this joint distribution is $1/\lambda_i+1/\xi_i$. From this, one realizes that the terms in brackets in the denominators of (\ref{eq:rho}) and (\ref{eq:chi}) correspond to the average durations of employment plus unemployment of agents at firm $i$. The factor $h_i\langle h\rangle_{\Gamma_i}k_i$ corresponds to the probability to enter and exit each edge connected to $i$. Therefore, $\rho$ measures the amount of overall job mobility in the entire economy, and $\chi$ corresponds to the per-agent $\rho$.

Expressions (\ref{eq:Lh}) and (\ref{eq:Uh-simpler}) are very similar, the only difference being the exchange of $\lambda_i$ and $\xi^{(h)}_i$. By taking the quotient of (\ref{eq:Lh}) and (\ref{eq:Uh-simpler}) we obtain
\begin{equation}
\frac{\langle L_i\rangle^{(v)}}{\langle U_i\rangle^{(v)}}=\frac{\xi^{(v)}_i}{\lambda_i}.
\label{eq:LoUvsXioLamb}
\end{equation}
In Sec.~\ref{sec:empirical-illustration}, we compare this result with empirical data (see Eq.~(\ref{eq:LoU-data})). This relation is potentially useful because (\ref{eq:LoUvsXioLamb}) only involves model quantities that can be measured in the empirical data. Note that the ratio in (\ref{eq:LoUvsXioLamb}) measures how much time an agent spends employed at firm $i$ compared to the time the agent spends looking for a job among the neighbors of $i$. 
\subsection{Application of the model}
\label{sec:algorithm}
An attempt to apply our model to real-world situations runs into the difficulty that data is not available for all parameters. In particular, we do not have data to determine the rates of opening of positions $\{v_i\}$ nor hiring rates $\{h_i\}$ of firms.
These parameters, relevant from the economic standpoint as they allow calculation of endogenous effects for each firm, are not usually collected by statistical authorities. These difficulties, however, can be overcome by expressing the equations of the system in terms of available information.

As observed above in (\ref{eq:lambda-flows}), $\boldsymbol{\Lambda}$ can be written in terms of outflows which can be directly measured from data, and then inserted into (\ref{eq:LambdaEq}). We also require a method to express the uniqueness condition (\ref{eq:cond}) in terms of the data. Note that (\ref{eq:cond}), with the use of (\ref{eq:Uavg}) and the definition of $\xi_i$, can be rewritten as
\begin{equation}
H=\sum_{i=1}^N(\langle L_i\rangle+\langle U_i\rangle)
=\sum_{i=1}^N\left[\frac{1}{\lambda_i}+\frac{1}{\sum_{\gamma_j\neq\emptyset}\langle h\rangle_{\gamma_j}{\rm Pr}(\gamma_j)}\right]\lambda_i\langle L_i\rangle
=\sum_{i=1}^N\left(\frac{1}{\lambda_i}+\frac{1}{\xi_i}\right)\lambda_i\langle L_i\rangle
\label{eq:cond-emp}
\end{equation}
where both $\lambda_i$ and $\xi_i$ can be measured via average employment
and unemployment times of workers at firm $i$.

Using (\ref{eq:cond-emp}), one can construct a modified matrix $\boldsymbol{\tilde{\Lambda}}$ where one of the rows from $\boldsymbol{\Lambda}$ is eliminated (any row) and substituted by a row based on (\ref{eq:cond}). This leads to a non-homogeneous linear set of equations with a unique solution. One possible concrete form of this can be to eliminate the last row to generate $\boldsymbol{\tilde{\Lambda}}_{ij}$ with the form
\begin{equation}
\boldsymbol{\tilde{\Lambda}}_{ij}:=
\left\{
\begin{array}{lcl}
\boldsymbol{\Lambda}_{ij}&\qquad&[1\leq i\leq N-1]\\
\frac{1}{\lambda_j}+\frac{1}{\xi_j}
&\qquad&[i= N].
\end{array}
\right.
\end{equation}
With $\boldsymbol{\tilde{\Lambda}}$ defined in this way, one can further introduce
\begin{equation}
\mathbf{Y}_j:=
\left\{
\begin{array}{lcl}
0&\qquad&[1\leq j\leq N-1]\\
H&\qquad&[j=N]
\end{array}
\right.
\end{equation}
leading to the matrix equation
\begin{equation}
\boldsymbol{\tilde{\Lambda}}\mathbf{X}=\mathbf{Y},
\label{eq:LambdaTildeEq}
\end{equation}
where $\mathbf{X}$ is still the column matrix of employment sizes as in (\ref{eq:LambdaEq}).

As a final ingredient, we relax the condition that $G$ is a connected graph, and allow for the presence of $C$ disconnected components. We continue to assume that the structure of the components is generally non-trivial in real data, which is to say that in what follows we do not expand on the cases where $G$ has isolated nodes or very small components (containing, say, only two nodes) where the behavior is trivially simple but requires more technicalities to be described with rigor. 

It is known that for a graph with $C$ connected components, the rank of the adjacency matrix is $N-C$. This reflects the fact that the dynamics of walkers on each component runs independently of the other components due to the lack of connections among them. This reduced rank value corresponds to the need for $C$ distinct conditions stemming from the number of workers on each isolated component. For a given initial distribution of such workers $\{H_c\}_{c=1,\dots,C}$ where $\sum_{c=1}^C H_c=H$ over the components, one obtains a set of $C$ conditions
\begin{equation}
\begin{array}{lcl}
\sum_{i\in G_1}(\langle L_i\rangle+\langle U_i\rangle)&=&H_1\\
&\vdots&\\
\sum_{i\in G_C}(\langle L_i\rangle+\langle U_i\rangle)&=&H_C
\end{array}
\end{equation}
where $G_c$ is the $c$ component among $C$. The modified matrix $\boldsymbol{\tilde{\Lambda}}$ can now be constructed in a similar way as for the single component case. As a prior step to the construction, we relabel the nodes so that the adjacency matrix becomes block diagonal, with each block corresponding to the adjacency matrix of a single connected component. In this way, we can write for each component fundamentally the same equation (\ref{eq:LambdaTildeEq}) now indexed by $c$. We can also write a single matrix equation, where the block diagonal shape of $\mathbf{A}$ has been introduced, and $\boldsymbol{\tilde{\Lambda}}$ takes the form 
\begin{equation}
\boldsymbol{\tilde{\Lambda}}_{ij}:=
\left\{
\begin{array}{lcl}
\boldsymbol{\Lambda}_{ij}&\qquad&[1\leq i\leq S_1-1]\\
\frac{1}{\lambda_j}+\frac{1}{\xi_j}&\qquad&[i= S_1]\\
&\vdots&\\
\boldsymbol{\Lambda}_{ij}&\qquad&[S_{C-1}\leq i\leq S_C-1]\\
\frac{1}{\lambda_j}+\frac{1}{\xi_j}&\qquad&[i= S_C]
\end{array}
\right.
\end{equation}
with
\begin{equation}
\mathbf{Y}_j:=
\left\{
\begin{array}{lcl}
0&\qquad&[1\leq j\leq S_1-1]\\
H_1&\qquad&[j=S_1]\\
&\vdots&\\
0&\qquad&[S_{C-1}\leq j\leq S_C-1]\\
H_C&\qquad&[j=S_C]
\end{array}
\right.
\end{equation}
and $\mathbf{X}_i=\lambda_i\langle L_i\rangle$, as before.

The previous comments regarding the number of connected components can be related to the dynamics of the individual agent treated in Sec.~\ref{sec:agent}. It is clear that, in the case of $C$ components, a prerequisite to analyzing the problem is to provide an initial condition that specifies the location of the agent. If the agent is placed at component $c$ at time $t=0$, solving for $r_\infty(i)$ and $s_\infty(i)$ provides information relevant to $c$ only. If, on the other hand, we specify that at time $t=0$ the agent can be found in component $c$ with a probability $H_c/H$, then we can develop analysis to describe the more general case of agents distributed across the graph. 

In this section, we have developed one particular approach for solving (\ref{eq:LambdaEq}), but other approaches are possible. Among those, one of the most general is to apply singular value decomposition to determine a decomposition for the space of solutions of $\boldsymbol{\Lambda}$, including the kernel of the matrix, and then apply the uniqueness conditions. Ultimately, the application of a particular method is a practical matter that takes into consider aspects of the problem that may go beyond the theoretical ones.
\section{Empirical analysis}
\label{sec:empirics}
The theory developed above rests on a set of assumptions described in
the previous section; namely, the presence of a well defined network, a
steady state flow of workers during significant periods of time, and a 
simple approximation to the possible behavior of workers as they leave 
firms and seek new employment. The aim of this section is to explore 
empirically the model, both by corroborating that 
the assumptions we make are close to the observed behavior of the 
system, and by studying some of the consequences of our model in terms 
of how well it predicts the statistical characteristics of real 
systems. We find that both our assumptions and the results that the 
model predicts fit the data well.

We now describe our approach in more detail. The first assumption employed is that the edges of the graph can reasonably be considered static. We test this by introducing the notion of {\it persistence} of the graph, i.e., the property that over successive time intervals, many edges produced by agent's job transitions between firms re-occur, indicating that the graph does not change over time in a random way, but instead exhibits a static behavior (at least partially). By restricting our model to static networks, our approach assumes labor markets are static, and although this is not entirely true, it appears that it is sufficiently true to capture the main behavior of the system as our analysis indicates\footnote{The accuracy of this assumption is affected by time scales, as over very long periods of, say, a decade or more, one could not expect static behavior. But this is acceptable in our framework, as one cannot expect to predict the full economy over such time scales anyway.}. 

The second assumption tested is that the system is close to the steady state. To confirm this assumption, we study the histograms of agent in and out flows at firms, finding that, by far, the most typical situation is that these flows are virtually balanced for each firm, indicating that in fact the system is typically not growing or declining, but rather remaining steady on a firm by firm basis.

In order to characterize the agreement between empirical data and our model, we also measure the per-firm values of the separation rate $\lambda_i$ and the job finding rate $\xi_i$. These two quantities, which we assume to be related to the firms rather than the individual agents, appear to be satisfactory quantities when comparing model and data.

As a way to illustrate the fact that our framework captures some of the relevant features of job mobility, we test the consequences of the homogeneous opening rates model ($v_i=v$) against data. In particular, we study the ratio $\langle L_i\rangle$ and $\langle U_i\rangle$, vs. the ratio of $\lambda_i$ and $\xi_i$ (in effect checking eq.~(\ref{eq:LoUvsXioLamb})), and also whether $\langle L_i\rangle$ and $\langle U_i\rangle$ are consistent with the predicitions of (\ref{eq:Lh}) and (\ref{eq:Uh-simpler}). We find that indeed the model and data agree sufficiently to accept our approach as a plausible way to model job mobility.

We begin our detailed analysis with a description of the data we use, and then proceed to present the tests mentioned above.
\subsection{The data}
We use a high-resolution dataset and a support dataset where the information is more aggregated. The main dataset consists of employer-employee matched records at a daily resolution. It is a sample of $\approx 4\times 10^{5}$ workers and $\approx 8\times 10^4$ firms provided by the ``Instituto Mexicano del Seguro Social" (Mexican Social Security Institute or IMSS). The workers were sampled from the universe of individuals who were registered at IMSS between 2000 and 2008 (all individuals who work in the private sector are registered at IMSS). Then, the complete employment history of each worker was extracted from the database (that includes any activity before 2000). The fraction of workers captured in this dataset is approximately 1\% of the total workforce of Mexico in the private sector.

These employer-employee matched records are constructed in the following way. For each worker, every time there is a job transition between employment and unemployment (in either direction), the worker's record is updated: $i$) when hired into a firm, the record contains the day at which the worker starts employment and a unique identifier for the firm (consistent for all workers in the data set), and $ii$) when separated from a firm, the day in which this occurred. Note that the dataset does not track firms directly, and thus the only means of tracking them is through individuals in the dataset. 

We employ the IMSS dataset for our main analysis due to its high resolution. The support dataset consists of employer-employee matched records from the universe of employed workers and firms in Finland, constructed from social security records, and provided by Statistics Finland. These records consist of annual observations that track each employed worker in the economy between 2000 and 2008. Each year contains approximately 200,000 firms and 1.5 million individuals. 

\subsection{Persistent flows}
\label{sec:persistence}
There is some empirical evidence that, whenever a person leaves firm $i$ and then gets a job at firm $j$, transitions between $i$ and $j$ (in either direction) are likely to be repeated in the future~\cite{Collet}. If indeed such transitions are repeated, we consider them to be \emph{persistent}. We employ our datasets in order to measure persistence in both Mexico and Finland. 

Let $t$ and $t+1$ be the starting and ending times of a given period of job transitions, $t+1$ to $t+2$ a second period, etc.\footnote{We concentrate on annual consecutive periods. However, non consecutive periods yield similar results.} We denote such time intervals with $I(t;1):=[t,t+1]$ and use $I(t):=I(t;1)$ for simplicity. To construct a graph of firms and transitions based on empirical data, we proceed as follows: when we observe a worker transitioning from firm $i$ to $j$ or vice versa, we introduce an undirected edge between $i$ and $j$; we do not consider weights, so once an edge has been created, additional $i\leftrightarrow j$ transitions in the same period have no further consequences in the graph structure. In addition, our data allows us to observe firms that may not have had any incoming or outgoing job transitions, and these firms are encoded as isolated nodes. The graph $G_{I(t)}$ is constituted by all the edges occurring in the period $I(t)$, the nodes to which those edges are incident, and the isolated nodes that display no transitions. To measure persistence, the relevant question is: how many edges in period $I(t+1)$ also occurred in the previous period $I(t)$? To assess this, we define
\begin{equation}
PE_{t}=E(G_{I(t)})\cap E(G_{I(t+1)}),
\label{eq:edge-overlap}
\end{equation}
the set of common edges between graphs $G_{I(t)}$ and $G_{I(t+1)}$, where $E(.)$
is set of edges of the argument graph.
Then, $|PE_{t}|/|E_{I(t+1)}|$ is the fraction of the edges $E(G_{I(t+1)})$ that are persistent. 

This concept of persistence captures repetition of job transitions. However, a random job search process can produce repeated job transitions by chance. Therefore, persistence is only meaningful to the extent that it occurs more frequently than what a random process would lead to. Furthermore, one should be able to define confidence intervals addressing whether the persistence found could emerge as a consequence of random fluctuations. A natural random (null) model one could use to compare persistence in real vs. random job search is to allow any individual looking for a job to apply and potentially fill any of the vacancies offered by the firms of the graph. In this model, firms have a defined number of vacancies and of job-seekers (both determined from our datasets) and individuals are allowed to apply and potentially be hired into any of those jobs (except the ones of its last employer). Below, we develop a set of statistical tests to determine confidence intervals for persistence using this approach, and apply it to the IMSS data from Mexico. Given that the absence of an edge is potentially meaningful because it may signal a genuine lack of affinity between firms that never connect, we perform additional confidence testing to take this into account. We find that there is a large degree of confidence that persistence is indeed present, and that adding tests to account for lack of connections only increases the confidence levels. We should briefly mention here that our null model is indeed an appropriate test to compare to current economic thinking, which assumes aggregate matching processes that ignore firms and their contributions into the heterogeneity of the real job transition process. 
\subsubsection{Hypothesis testing for persistence}
The null models are constructed independently for every pair of consecutive time intervals. For one such pair of time intervals, we first determine for each firm $i$ and time interval $I(t)$ or $I(t+1)$ the number of hires into $i$ coming from other firms ($\eta_{.i}(t)$ and $\eta_{.i}(t+1)$) and job separations ($\Delta_{u,i}(t)$ and $\Delta_{u,i}(t1)$) that occur. For each of the intervals, for instance $I(t)$, a random job transition graph $G^{(r)}_{I(t)}$ is built by taking for each node the number $\eta_{.i}(t)$ as vacancies that need to be filled by $i$, and $\Delta_{u,i}(t)$ as the number of individuals that leave $i$ and seek other jobs. With these two number as constraints for every node, the vacancies and job-seekers are then randomly matched over the entire set of nodes, forbidding job-seekers to go back to their previous employer. This approach is basically equivalent to a random configuration model (for a review, see~\cite{Newman_SIAM}). A number $M=300$ of such random realizations is computed for each interval $I(t)$, generating an ensemble of random graphs. We use this ensemble to obtain the distribution of the statistic 
\begin{equation}
\psi_{t} = \frac{|PE'_{t}|}{|PE_{t}|},
\label{eq:psi}
\end{equation}
where $PE'$ and $PE$ are defined via (\ref{eq:edge-overlap}), with $PE'$ representing the fraction of persistent edges between random graph samples, and $PE$ the fraction of persistent edges between the corresponding empirical graphs. There are ${M\choose 2}$ values of $PE'$ generated by our procedure. The statistic $\psi_{t}$ measures the extent to which the global random matching mechanism explains the observed transitions over the ensemble built for multiple pairs of years covering an overall span of 8 years.

As mentioned above, the absence of an edge can contain relevant information about the lack of affinity between pairs of nodes. In the economics literature this would be thought of as a {\it friction}. Therefore, if the global search model explains both the observed persistence as well as the persistent lack of labor flows between firms, one would require an additional statistic to capture the persistence of lack of connections. We therefore define
\begin{equation}
\varrho_{t} = \frac{|PF'_{t}|}{|PF_{t}|},
\label{eq:varrho}
\end{equation}
where $PF$ denotes the set of persistent frictions (the pairs of firms that are not connected in the network).

\begin{center}
\begin{figure}
\centering
\includegraphics[scale=.5]{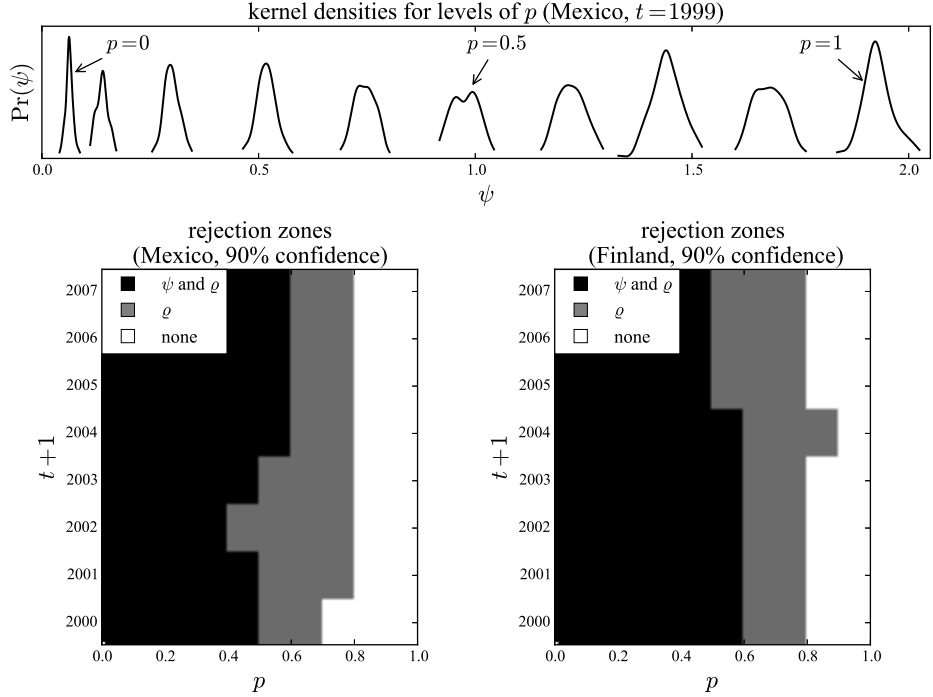}
\caption{Measures of persistence for job transitions. Top: Distributions of $\psi$ from (\ref{eq:psi}) when applied in the context of Sec.~\ref{sec:tunable-p} for various $q$. These distributions were generated from a Monte Carlo procedure. As the probability of local search $q$ increases, the distribution of $\psi$ shifts to the right. When $\psi$ falls inside the distribution (e.g., bellow the $90^{th}$ percentile) it means that the corresponding level of $q$ is enough to not reject the null model. This level is approximately $q=0.5$ for $\psi$ and $q=0.8$ for $\varrho$. Bottom: Rejection zones. The panels show the levels of $q$ for which the null hypothesis is rejected. The black area represents the case in which the null model is rejected for both $\psi$ and $\varrho$. For a $q$ in the gray zone, the null model is rejected only for $\psi$. In the white area the null model explains the empirical levels of both $\psi$ and $\varrho$. Synthetic distributions were created for values of $q \in [0,1]$ equally spaced by 0.1.}
\label{fig:persistence}
\end{figure}
\end{center}

Using Monte Carlo simulation, we performed a one-sided test for $\psi$ and $\varrho$. The null hypothesis is that, under a global search process, we would expect $\psi=1$ and $\varrho=1$. The global search hypothesis was rejected with 99\% confidence in both cases. For an illustration, the top panel in figure \ref{fig:persistence} shows the probability distribution of $\psi$ (the one on the far left) generated form the Monte Carlo procedure. Clearly, the confidence interval of the distribution is far bellow 1 (the mean is $\psi=0.001$), implying that the global search fails to explain the persistent labor flows between firms.
\subsubsection{A tunable model for the contribution of persistence}
\label{sec:tunable-p}
The global search mechanism fails to explain both the persistent edges and frictions across the eight years of data. This is consistent with intuition, given the large space of possible matches that can emerge when all job vacancies are accessible to all job-seekers. If in reality, as our results suggest, job-seekers use a subset of possible job transitions, the matching mechanism of our null model should restrict the job search. Therefore, we introduce an additional mechanism: with probability $q$ a job seeker searches through the graph and with probability $1-q$ searches globally. Clearly, when $q \rightarrow 1$ we obtain the mechanism proposed in section \ref{sec:model} and when $q \rightarrow 0$, the search is global over all firms.

With the local search mechanism in place, we need an additional assumption for the null model. Consider the null networks $G_{I(t)}$ and $G_{I(t+1)}$ with corresponding sets of edges $E(G_{I(t)})$ and $E(G_{I(t+1)})$. Since we are in a steady state, it is reasonable to assume that any edge in $G_{I(t)}$ can also exist in $G_{I(t+1)}$ (and vice versa), even if it is not observed in the data. Then, when a worker searches locally under the null model, it does so by using the network $G_{t}^*$, such that $E_t^* = E(G_{I(t)}) \cup E(G_{I(t+1)})$. This assumption captures the time-invariant aspect of the steady state, and allows $\varrho$ to take values higher than 1.

We compute the null model for different levels of $q$, so we can answer the question: what is the minimum $q$ needed to generate at least the level of persistence observed in empirical data. First, we randomize the matches between job seekers and vacancies, generating new datasets. Then, we construct null networks from these datasets. Next, we compute \eqref{eq:psi} and \eqref{eq:varrho} for each pair of null networks to generate their distributions. Finally, we use these distributions to perform a one-sided test with for each statistic. If the statistic falls beyond the $90^{th}$ percentile, the null model does not explain the persistence of edges or frictions. When we find a $q$ such that the statistic is below the $90^{th}$ percentile we cannot reject the null model. The smallest $q$ under which we cannot reject the null model for neither $\psi$ nor $\varrho$ is an indicator of at least how frequently people should search locally in a model in order to explain the structure of the empirical data.

Figure \ref{fig:persistence} shows the results from this analysis. In general, a higher $q$ is needed to explain both persistent edges and frictions than just edges. An approximate estimate suggests, in order to explain empirical persistence, a job-seeker needs to search on the network at least 75\% of the time. This result is consistent across both datasets and strongly suggests that the network approach is much more empirically relevant than the global search one, providing a solid motivation for the model developed in this paper.
\begin{figure}
\begin{center}
\epsfig{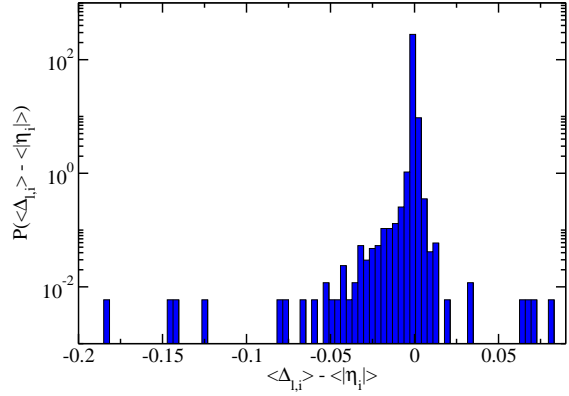}
\caption{Distribution of the difference between flow into and out of a firm. There are 49146
firms in this distribution.}
\label{fig:diff_flow}
\end{center}
\end{figure}
\subsection{Model validation}
\label{sec:model-val}
In this article we concentrate on the steady state behavior of the system. In order to validate this choice, we first study the distribution of $\langle\Delta_{l,i}\rangle-\langle|\boldsymbol{\eta}_i|\rangle$ from the data. From this point on, we concentrate on the IMSS data since its daily resolution allows us to identify the duration of employment and unemployment spells of each individual, which is crucial to our analysis. Then, if the system is close to the steady state, the distribution of $\langle\Delta_{l,i}\rangle-\langle|\boldsymbol{\eta}_i|\rangle$ should be concentrated around 0. Figure~\ref{fig:diff_flow} corresponds to the distribution of average agent daily flows over the period of 1 year into and out of a firm, with a pronounced peak around zero, which corresponds to our intuition. The averages have been taken by using the periods of observations of the workers associated with firms.

\begin{figure}
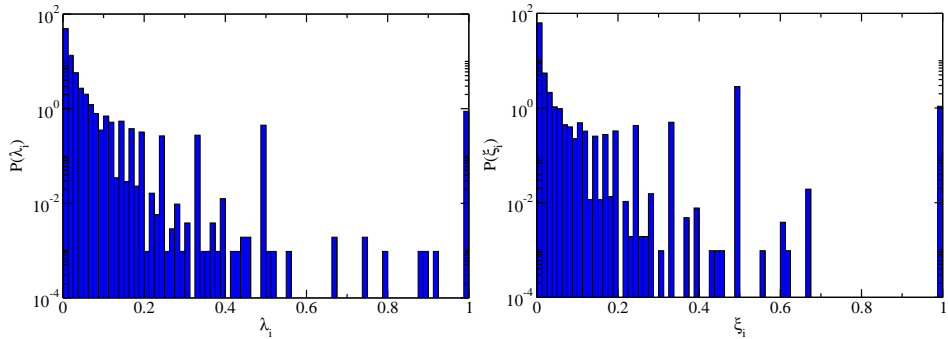

\epsfig{file=lambda_imss_2.eps,scale=0.25}
\epsfig{file=xi_imss_2.eps,scale=0.25}
\caption{(a) Distribution of values of $\hat{\lambda}_i$ over firms. In this case, $S^{(l)}=83450$.
(b) Distribution of $\hat{\xi}_i$ over firms, where $S^{(u)}=83450$.}
\label{fig:lambda_xi}
\end{figure}

Next, we determine the rates of separation and hiring. To estimate the values of separation rates, we proceed by tracking all employees of a firm that are observed to enter and exit that firm. Separation of an agent from a job is characterized by (\ref{eq:Lspell}). In order to estimate $\lambda_i$ for a firm, we perform a maximum likelihood estimation. For a sample of agents of size $S^{(l)}_i$, the log of the joint distribution of employment durations $\{t^{(l)}_1,\dots,t^{(l)}_{S^{(l)}_i}\}$ in a given firm is
\begin{equation}
\log \left[\prod_{j=1}^{S^{(l)}_i}{\rm Pr}(t^{(l)}_j)\right]=
\left(-S^{(l)}_i+\sum_{j=1}^{S^{(l)}_i}t^{(l)}_j\right)\log(1-\lambda_i)+S^{(l)}_i\log\lambda_i
\label{eq:log-like}
\end{equation}
and the maximum likelihood (ML) estimator is 
\begin{equation}
\hat{\lambda}_i:=\frac{S^{(l)}_i}{\sum_{j=1}^{S^{(l)}_i}t^{(l)}_j}=\frac{1}{\langle t^{(l)}_i\rangle},
\end{equation}
the value of $\lambda_i$ that maximizes (\ref{eq:log-like}). The effective rate of hiring $\xi_i$ can be estimated in the same way, with the ML estimator given by
\begin{equation}
\hat{\xi}_i:=\frac{S^{(u)}_i}{\sum_{j=1}^{S^{(u)}_i}t^{(u)}_j}=\frac{1}{\langle t^{(u)}_i\rangle}
\end{equation}
where there are $S^{(u)}_i$ unemployed individuals with unemployment times $\{t^{(u)}_1,\dots,t^{(u)}_{S^{(u)}_i}\}$. The measurements of $\hat{\lambda}_i$ and $\hat{\xi}_i$ can be studied via their distributions, as shown in Figs.~\ref{fig:lambda_xi}(a) and (b). For the distribution of $\hat{\lambda}_i$, the sample of individuals was restricted to those who began and ended their tenure of employment within the time frame of the data; similar considerations were applied to the distribution of $\hat{\xi}_i$, restricting the sample to individuals that become unemployed and subsequently found employment during the window of observation. The distributions of $\hat{\lambda}_i$ and $\hat{\xi}_i$ both exhibit decaying heavy-tails, indicating a wide variation in the rates of agent separation or hiring.
\subsection{An illustration: homogeneous opening rates}
\label{sec:empirical-illustration}
The analysis presented above supports a picture of considerable heterogeneity in real economic systems. Therefore, a full treatment of the data is likely to require detailed application of our model, accompanied by robust statistical analysis that is yet to be fully developed.

However, for the purposes of illustration in this article, it is useful to perform some basic comparisons between the data and some version of our model. Given the absence of information for $\{v_i\}$ and $\{h_i\}$, it seems reasonable to compare a model that simplifies at least one of these parameters while assuming the other continues to be heterogeneous. This provides some flexibility so that the model is able to cope with at least some level of complexity from the real data. Therefore, we chose to compare the data with the model characterized by homogeneous rates $v_i=v$ for firms to accept applications (opening rates). We find that, even for this simple case, there is evidence to support the plausibility of our approach.

As a first test, we explore the ratio (\ref{eq:LoUvsXioLamb}), which is convenient because it only contains directly measured parameters. Note that here, since all the parameters emerge from measurement, we are not concerned with using the superindex $(v)$ to symbolize the homogeneity in $v$. Using the dataset from Mexico, we estimated ${\langle L_i\rangle }$ and ${\langle U_i\rangle}$ for 2008. In order to assure independence across the errors, we estimate ${\hat{\xi}_i}$ and ${\hat{\lambda}_i}$ from observations of employment and unemployment spells that concluded at least three years prior to 2008. We excluded firms for which $\langle U_i\rangle=0$ and estimate $\hat{\alpha}$ and $\hat{\beta}$ defined by
\begin{equation}
\frac{\langle L_i\rangle }{\langle U_i\rangle} = \alpha \left(\frac{\hat{\xi}_i}{\hat{\lambda}_i}\right)^{\beta}.
\label{eq:LoU-data}
\end{equation}

Due to the large variance heterogeneity in the data, we make use of the random re-sample consensus algorithm (RANSAC) \cite{Fischler} in order to estimate $\alpha$ and $\beta$. The algorithm randomly samples the data in order to discriminate the outliers and fit \eqref{eq:LoU-data} via OLS to the in-liers iteratively. Since the RANSAC algorithm is non-deterministic, the estimators vary from run to run. In order to illustrate the coherence of the model, we performed 10,000 estimations using this procedure and analyzed the distribution of $\hat{\alpha}$ and $\hat{\beta}$.

\begin{center}
\begin{figure}
\centering
\includegraphics[scale=.6]{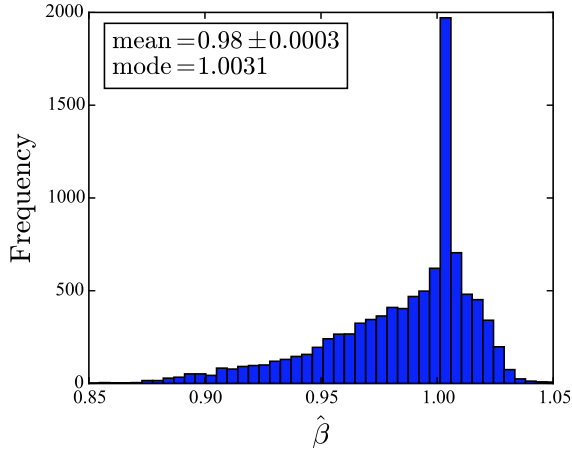}
\caption{Distribution of $\hat{\beta}$ obtained form 10,000 estimations of the RANSAC algorithm, using OLS as the underlying model.}
\label{fig:coefficient}
\end{figure}
\end{center}

Figure \ref{fig:coefficient} shows the histogram of the estimator $\hat{\beta}$. The average $\beta$ is 0.98, while the most frequent is 1.0031. The average estimator $\hat{\alpha}$ of the intercept is $1.1425 \pm 0.0007$. These  results are quite close to the theoretical prediction of (\ref{eq:LoUvsXioLamb}).

To perform a second test, we consider whether (\ref{eq:Lh}) and (\ref{eq:Uh-simpler}) may be consistent with the data. For this, we concentrate on two conditional probabilities: i) ${\rm Pr}(L_i|k_i/\hat{\lambda}_i)$ for the number of employed individuals at a firm, given the firm is characaterized by the ratio $k_i/\hat{\lambda}_i$, and ii) ${\rm Pr}(U_i|k_i/\hat{\xi}_i)$ for the number of unemployed individuals at a firm, given the firm is characterized by the ratio $k_i/\hat{\xi}_i$. In particular, we want to learn whether the basic predictions contained in (\ref{eq:Lh}) and (\ref{eq:Uh-simpler}) are satisfied, i.e., that $\langle L_i\rangle\sim k_i/\hat{\lambda}_i$ and $\langle U_i\rangle\sim k_i/\hat{\xi}_i$. 

\begin{figure}
\begin{center}
\epsfig{file=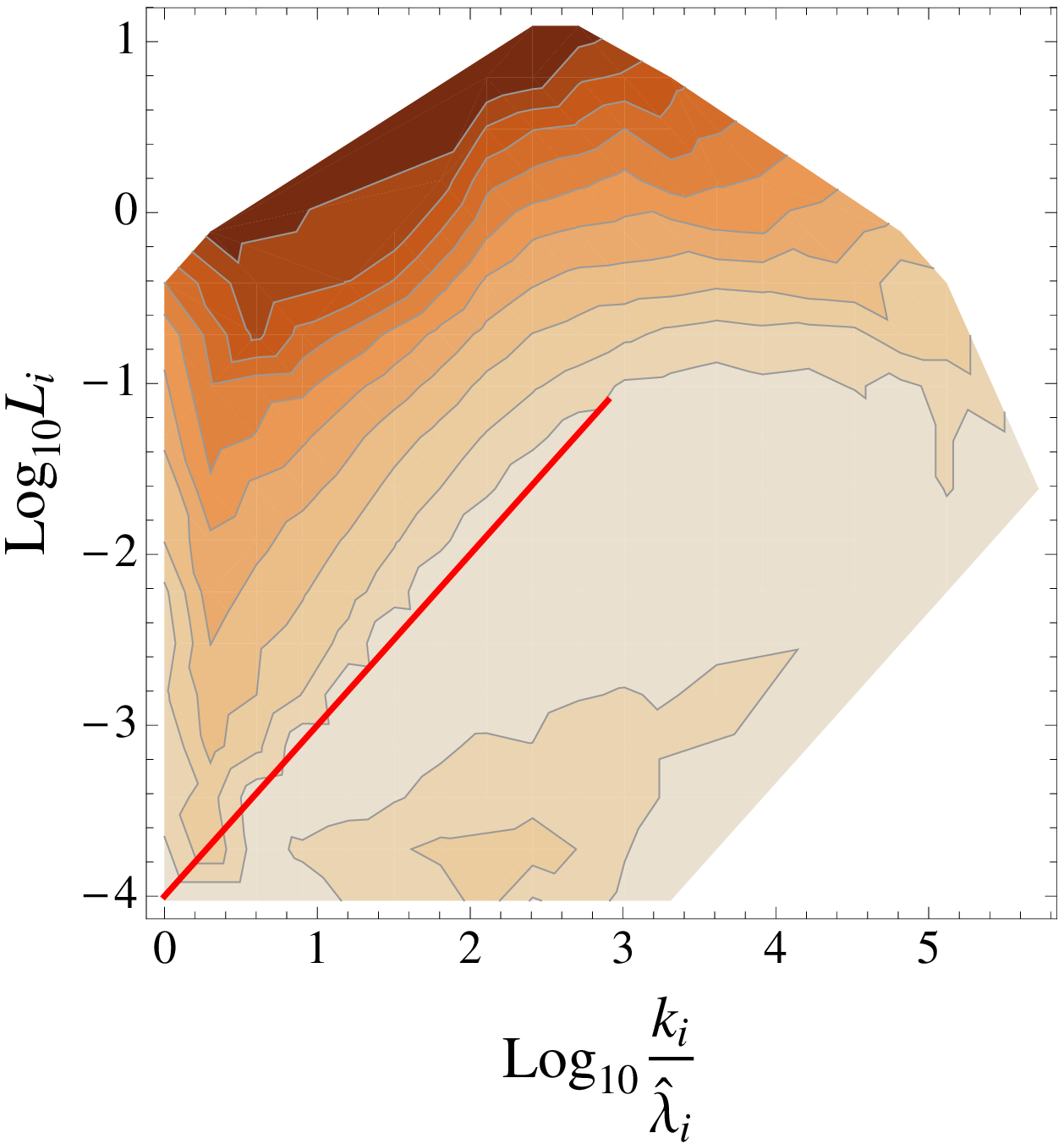,scale=0.38}
\epsfig{file=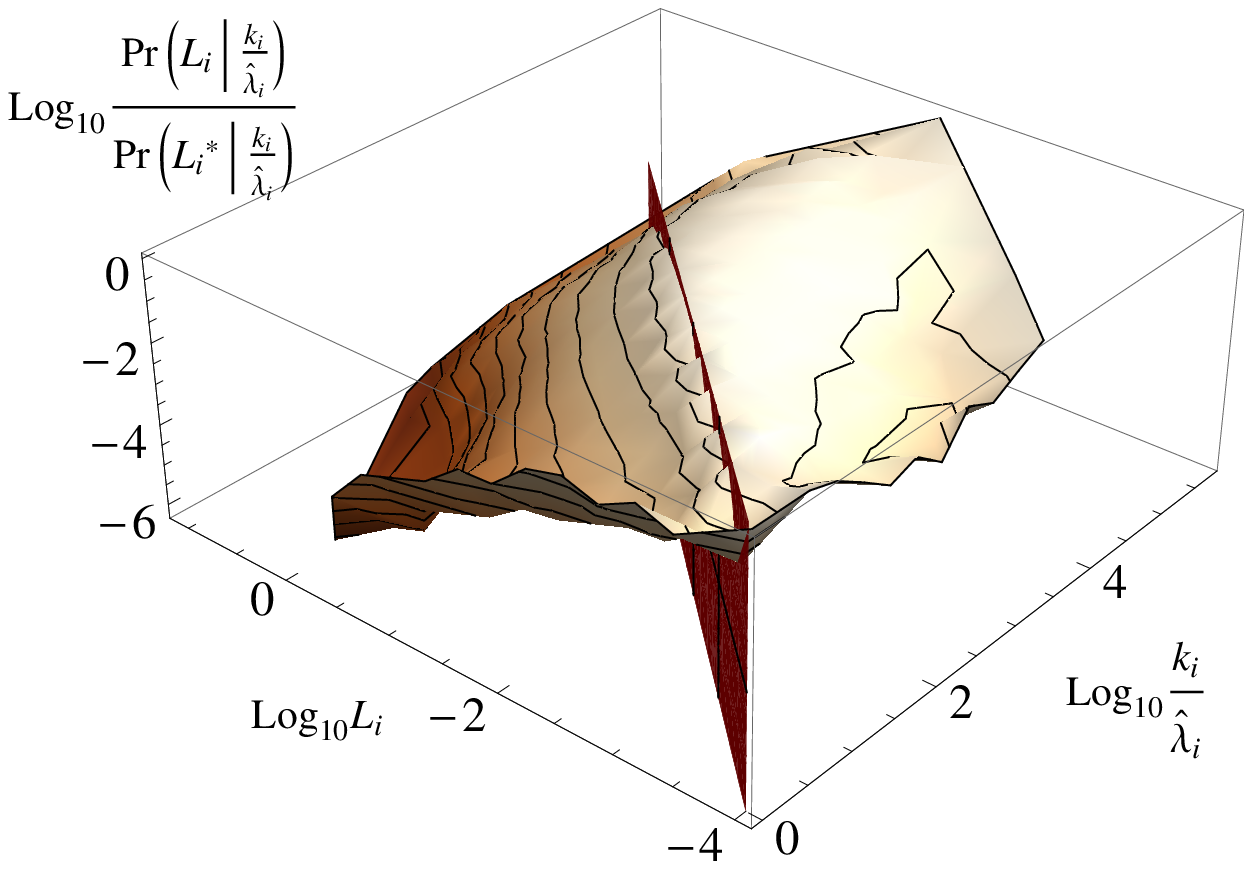,scale=0.62}
\caption{Behavior of $\log_{10}\left[{\rm Pr}(L_i|k_i/\hat{\lambda}_i)/{\rm Pr}(L^*_i|k_i/\hat{\lambda}_i)\right]$ from 49854 firms. (a) Contour plot together with a linear map $\alpha k_i/\hat{\lambda}_i$ with $\alpha$ a numerical constant. This line is close to parallel to the contour line of the maximum of the ratio of distributions, which indicates that the linear relationship $L^*_i\sim \alpha k_i/\hat{\lambda}_i$ is plausible for a range of values of $k_i/\hat{\lambda}_i$. (b) The corresponding 3-dimensional plot of the ratio of probabilities. The plane coincides with the line drawn in (a), defined by the parametric representation $(k_i/\hat{\lambda}_i,k_i/\hat{\lambda}_i,\beta)$ where $\beta$ is a free parameter.}
\label{fig:Lklamb}
\end{center}
\end{figure}

\begin{figure}
\begin{center}
\epsfig{file=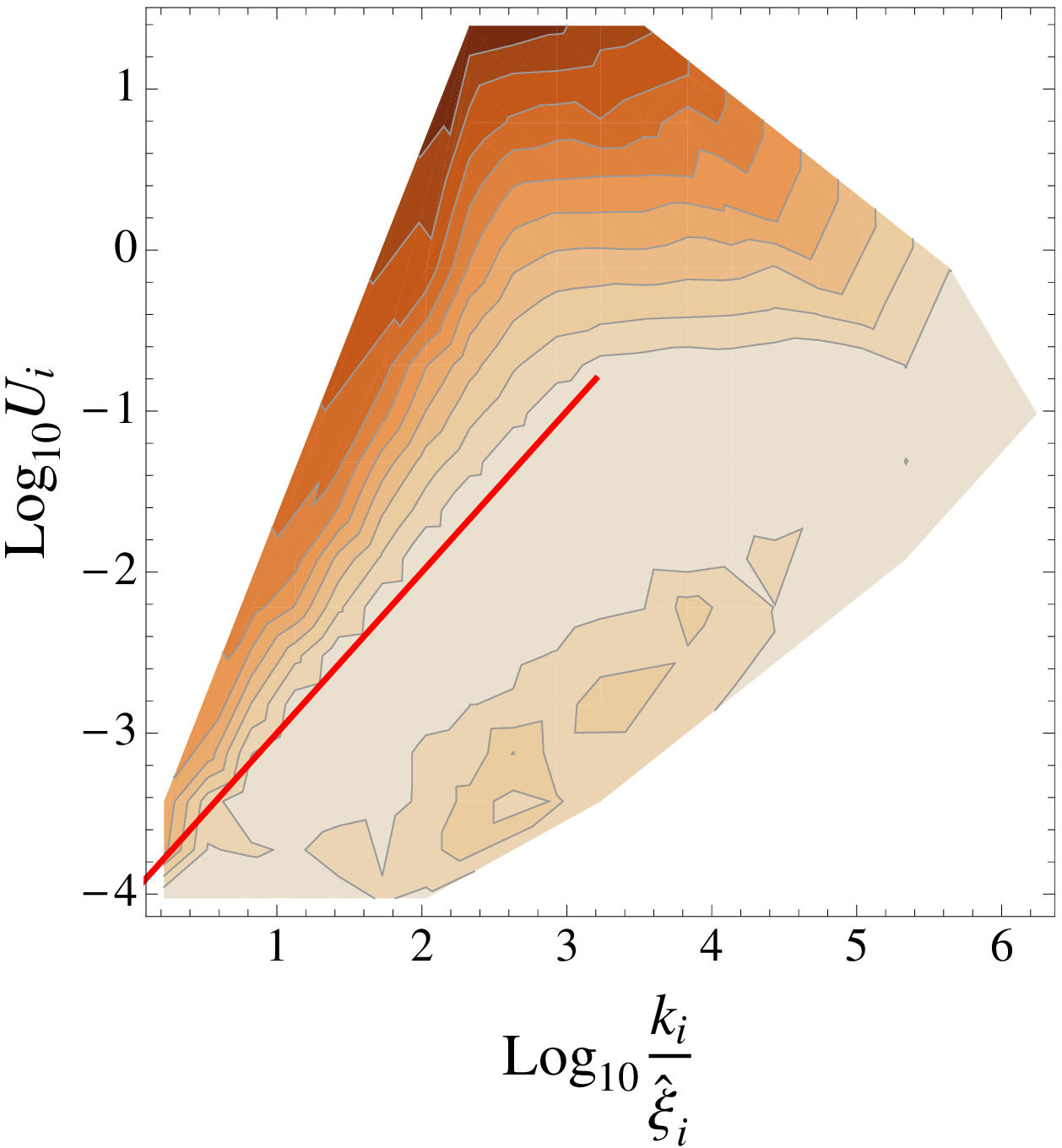,scale=0.38}
\epsfig{file=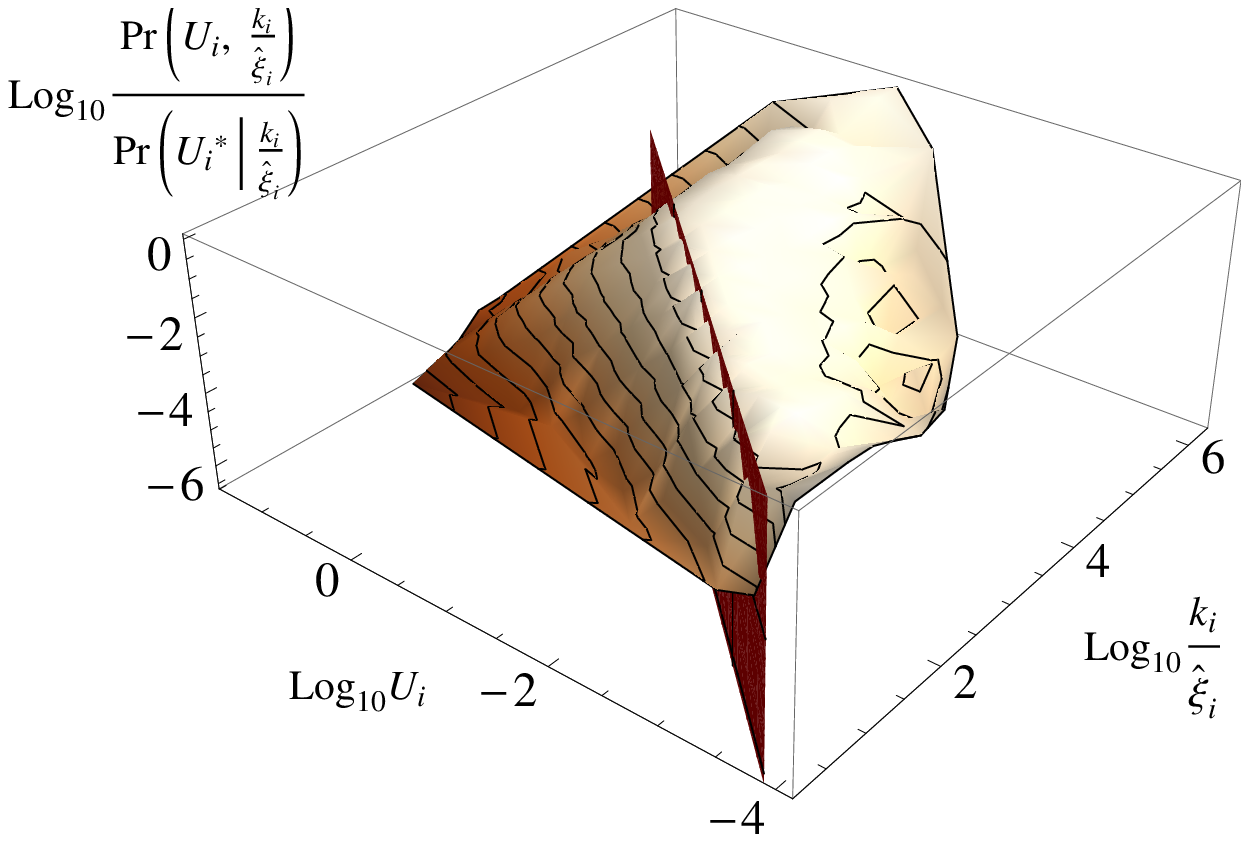,scale=0.62}
\caption{Behavior of $\log_{10}\left[{\rm Pr}(U_i|k_i/\hat{\xi}_i)/{\rm Pr}(U^*_i|k_i/\hat{\xi}_i)\right]$ from 49854 firms. The interpretation of this plot is analogous to that of Fig.~\ref{fig:Lklamb}. (a) Contour plot together with a linear map $\alpha k_i/\hat{\xi}_i$ with $\alpha$ a numerical constant. This line is close to parallel to the contour line of the maximum of the ratio of distributions, which indicates that the linear relationship $U^*_i\sim \alpha k_i/\hat{\xi}_i$ is plausible for a range of values of $k_i/\hat{\xi}_i$. (b) The corresponding 3-dimensional plot of the ratio of probabilities. The plane coincides with the line drawn in (a), defined by the parametric representation $(k_i/\hat{\xi}_i,k_i/\hat{\xi}_i,\beta)$ where $\beta$ is a free parameter.}
\label{fig:Ukxi}
\end{center}
\end{figure}

In Figs.~\ref{fig:Lklamb} and \ref{fig:Ukxi}, we present contour and 3-dimensional plots of $\log_{10}\left[{\rm Pr}(L_i|k_i/\hat{\lambda}_i)/{\rm Pr}(L_i^*|k_i/\hat{\lambda}_i)\right]$, and $\log_{10}\left[{\rm Pr}(U_i|k_i/\hat{\xi}_i)/{\rm Pr}(U_i^*|k_i/\hat{\xi}_i)\right]$, respectively. Here, ${\rm Pr}(L_i^*|k_i/\hat{\lambda}_i)$ and ${\rm Pr}(U_i^*|k_i/\hat{\xi}_i)$ correspond to the probabilities associated with the conditional modes of $L_i$ and $U_i$. The reason to plot the ratios just defined is that (\ref{eq:Lh}) and (\ref{eq:Uh-simpler}) are concerned with averages rather than distributions, and we therefore must devise a way to relate the empirical analysis with our predictions. To interpret the plots, we introduce a line of slope 1 (linear relation) in Figs.~\ref{fig:Lklamb}(a) and \ref{fig:Ukxi}(a). Such lines, by definition, scale as $k_i/\hat{\lambda}_i$ and $k_i/\hat{\xi}_i$. The relevant feature that the plots show is that these lines runs parallel to the contour for the largest value of ${\rm Pr}(L_i|k_i/\hat{\lambda}_i)$ and ${\rm Pr}(U_i|k_i/\hat{\xi}_i)$, or in other words, $L_i^*\sim k_i/\hat{\lambda}_i$ and $U_i^*\sim k_i/\hat{\xi}_i$. Figures~\ref{fig:Lklamb}(b) and \ref{fig:Ukxi}(b), showing in 3-dimensions the surface of the logarithm of the distribution ratios, reinforce our interpretation: in these plots the location of the maxima for the surfaces is cut by the planes that have been constructed to coincide with the linear maps of Figs.~\ref{fig:Lklamb}(a) and \ref{fig:Ukxi}(a). These relations hold for small and intermediate values of $k_i/\hat{\lambda}_i$ and $k_i/\hat{\xi}_i$, but eventually fail for the largest values of both ratios, probably due to poorer sampling at such large values of $k_i/\hat{\lambda}_i$ and $k_i/\hat{\xi}_i$. 

The results presented in this section focus on a very simple comparison and, notwithstanding the partial differences we encounter between our equations and the measurements, support the plausibility of our model as a way to explain job mobility. 
\section{Conclusion} 
\label{sec:conclusions}
Detailed high resolution data on employment at large scale is becoming rapidly available, and this provides an opportunity to revisit the way in which job mobility and labor flows are studied. In particular, it makes it possible to move away from aggregate models that, while having been very useful, have been unable to address some important outstanding problems, such as the construction of realistic shock scenarios, which are necessary if one is to attempt to design real-time forecasting models of high resolution employment flow. This task, which has not yet been possible, may be within our reach for the first time, with considerable potential value for economic policy design that is well grounded empirically and for which impacts can be forecast in great detail.

In this manuscript, we have introduced a new basic framework that takes into account the role of firms in employment, and makes extensive use of real data. By performing a number of tests, we have been able to see that indeed the model behaves in similar ways to the data. Furthermore, we have provided the basic ingredients for algorithms to calculate the average numbers of employed and unemployed agents associated with a firm. The notion of firm specific unemployment, which we have introduced here, is a new concept that allows us to keep track of the information that is implicitly contained in the fact that an agent has held a job in a certain firm, indicating that agent's affinity to some firms but not others of the economy.

An interesting consequence arising from (\ref{eq:LoUvsXioLamb}) is that in the steady state the numbers of employed and unemployed agents of a firm are not independent of each other and therefore, firms that have large numbers of employees could contribute large numbers of unemployed people if the ratio between the average times of employment and post-employment search is low. This is a question worth further exploration.

Finally, our introduction of a framework based on random walks on graphs to study job mobility can be a useful development. Random walks on graphs have a considerably long history, and a great deal is known about them (see, e.g.~\cite{Lovasz} for a review). Being able to deploy such a toolkit on questions regarding employment may lead to new results with potential academic and practical impact.

\section{Acknowledgements}
We thank Andrew Elliott, Jos\'e Javier Ramasco, Felix Reed-Tsochas, Gesine Reinert, Jari Saram\"aki, and Margaret Stevens for helpful discussions about the research and manuscript.


\begin{thebibliography}{1}
\bibitem{Petrongolo} B. Petrongolo and C. Pissarides,  
{\it Looking into the Black Box: A Survey of the Matching Function}, Journal of Economic Literature, 39 (2001), pp. 390-431.
\bibitem{Rogerson} R. Rogerson, R. Shimer, and R. Wright,
{\it Search-Theoretic Models Of The Labor Market: A Survey}, J. of Econ. Lit., 43 (2005) pp. 959-988.
\bibitem{Mortensen} D. Mortensen and C. Pissarides, {\it Job creation and job destruction in the theory of unemployment}, Rev. of Econ. Stud., 61 (1994), pp. 397–415.
\bibitem{Shimer} R. Shimer, {\it Mismatch}, Am. Econ. Rev., 97 (2007), pp. 1074-1101.
\bibitem{Stevens} M. Stevens, {\it New Microfoundations for the Aggregate Matching Function}, Int. Econ. Rev., 48 (2007), pp. 847-868.
\bibitem{Gabaix} X. Gabaix, {\it The Granular Origins of Aggregate Fluctuations}, Econometrica, 79 (2011), pp. 733-772.
\bibitem{diGiovanni} J. di Giovanni, A. Levchenko, I. and Mejean, {\it Firms, Destinations, and Aggregate Fluctuations}, Econometrica. 82(4) (2014), pp. 1303–1340.
\bibitem{Hamermesh} D. Hamermesh, {\it Fun with Matched Firm-Employee Data: Progress and Road Maps}, Labour Economics, 15 (2008), pp. 662–672.
\bibitem{Grimmett} G. Grimmett, {\it Probability on Graphs: Random Processes on Graphs and Lattices}, Cambridge University Press, Cambridge, United Kingdom, 2010.
\bibitem{Lovasz} L. Lov\'asz, {\it Random Walks on Graphs: A Survey}, Combinatorics, Paul Erd\"os is Eighty (Volume 2), Keszthely (Hungary), 1993, pp. 1–46.
\bibitem{Aldous} D. Aldous and J. Fill, {\it Reversible Markov chains and random walks on graphs (in preparation)}. Online version available at http://www.
stat.berkeley.edu/users/aldous/RWG/book.html.
\bibitem{Boorman} S. Boorman, {\it A Combinatiorial Optimization Model for Transmission of Job Information Through Contact Networks}, The Bell Journal of Economics, 6 (1975), pp. 216-249.
\bibitem{Montgomery} J. Montgomery {\it Social Networks and Labor-Market Outcomes: Toward an Economic Analysis}, Am. Econ. Rev., 81 (1991), pp. 1408-1418.
\bibitem{Calvo} A. Calv\'{o}-Armengol and M. Jackson, {\it The Effects of Social Networks on Employment and Inequality}, Am. Econ. Rev., 94 (2004), pp. 426-454. 
\bibitem{Schmutte} I. M. Schmutte, {\it Free to Move? A Network Approach to the Analysis of Job Mobility}, Lab. Econ. 29 (2014), pp. 49-61.
\bibitem{Fagiolo} F. Schweitzer et al. {\it Economic Networks: The New Challenges}, Science, 325 (2009), pp.422.
\bibitem{Granovetter} M. Granovetter {\it The Strength of Weak Ties}, American Journal of Sociology, 78 (1973), pp. 1360-1380.
\bibitem{Rodrigo} A. Rodrigo et al. {\it Markovian Networks in Labour Flows}, J. Op. Res. Soc. 57, (2006), pp. 526-531.
\bibitem{Guerrero} O. Guerrero and R. L. Axtell, {\it Employment Growth through Labor Flow Networks}, PLoS ONE, 8 (2013), e60808. doi:10.1371/journal.pone.0060808.
\bibitem{AGL} R. L. Axtell, O. Guerrero, and E. L\'opez, {\it The Network Composition of Aggregate Unemployment}, in preparation (2015). 
\bibitem{Bollobas} B. Bollob\'as, {\it Modern Graph Theory}, Springer, NY, 2002.
\bibitem{Wilf} H. S. Wilf, {\it Generatingfunctionology, 3rd ed.}, A. K. Peters/CRC Press, Wellesley, MA, 2006.
\bibitem{Hoffman} K. Hoffman and R. Kunze, {\it Linear Algebra, 2nd ed.}, Prentice-Hall, Upper Saddle River, NJ, 1971.
\bibitem{Collet} F. Collet F. and P. Hedstr\"omb, {\it Old friends and new acquaintances: Tie formation mechanisms in an interorganizational network generated by employee mobility}, Soc. Net. 35 (2013), pp. 288–299.
\bibitem{Fischler} M. Fischler and R. Bolles, {\it  Random Sample Consensus: A Paradigm for Model Fitting with Applications to Image Analysis and Automated Cartography}, Comm. of the ACM, 24, (1981), pp. 381–395.
\bibitem{Newman_SIAM} M. E. J. Newman, {\it The structure and function of Complex Networks}, SIAM Review 45, (2003) pp 167-256.
\end{thebibliography}
\end{document}